\documentclass{IEEEojcsys}
\makeatletter
\def\ps@pprintTitle{%
	\def\@oddhead{}%
	\def\@evenhead{}%
	\def\@oddfoot{}%
	\def\@evenfoot{}%
}
\makeatother 
\usepackage[colorlinks,urlcolor=blue,linkcolor=blue,citecolor=blue]{hyperref}

\usepackage{color,array}

\usepackage{graphicx}
\usepackage{cite}
\usepackage{amsmath,amssymb,amsfonts}
\usepackage{graphicx}
\usepackage{hyperref}
\hypersetup{hidelinks=true}
\usepackage{textcomp}
\usepackage{tikz}
\usepackage[english]{babel}
\usepackage{algorithm}
\usepackage[noend]{algpseudocode}
\usepackage{mathtools}
\usetikzlibrary{tikzmark}
\usetikzlibrary{calc}
\usepackage{pgfplots}
\usepackage{amsthm}
\usepackage{needspace}
\usepackage{booktabs}

\newtheorem{thm}{Theorem}

\newtheorem{lem}{Lemma}
\newtheorem{prop}{Proposition}

\newtheorem{rem}{Remark}

\newtheorem{defi}{Definition}
\newtheorem{ass}{Assumption}

\definecolor{rev}{rgb}{0,0,0}
\definecolor{new}{rgb}{0,0,0}

\definecolor{resl}{rgb}{0,0,0}
\definecolor{fin}{rgb}{0,0,0}
\newlength\figurewidth

\makeatletter
\newcommand\fs@spaceruled{\def\@fs@cfont{\bfseries}\let\@fs@capt\floatc@ruled
	\def\@fs@pre{\vspace{0.55\baselineskip}\hrule height.8pt depth0pt \kern2pt}%
	\def\@fs@post{\kern2pt\hrule\relax}%
	\def\@fs@mid{\kern2pt\hrule\kern2pt}%
	\let\@fs@iftopcapt\iftrue}
\makeatother

\begin{document}


\title{Robust stability of event-triggered nonlinear moving horizon estimation} 


\author{Isabelle Krauss\affilmark{1}}

\author{Victor G. Lopez\affilmark{1}  (Member, IEEE)}

\author{Matthias A. Müller\affilmark{1}  (Senior Member, IEEE)}

\affil{Leibniz University Hannover, Institute of Automatic Control,  30167 Hannover,  Germany} 

\corresp{CORRESPONDING AUTHOR: Isabelle Krauss (e-mail: \href{mailto:krauss@irt.uni-hannover.de}{krauss@irt.uni-hannover.de})}
\authornote{This work received funding from the European Research Council (ERC) under the European Union’s Horizon 2020 research and innovation programme (grant agreement No 948679).}


%
%
%

\setlength\figurewidth{0.8\columnwidth}

\begin{abstract}
	In this work, we propose an event-triggered moving horizon estimation (ET-MHE) scheme for the remote state estimation of general nonlinear systems. \textcolor{rev}{In the presented method, whenever an event is triggered, a single measurement is transmitted and the nonlinear MHE optimization problem is subsequently solved.}
If no event is triggered, the current state estimate is updated using an open-loop prediction based on the system dynamics.
Moreover, we introduce a novel event-triggering rule under which we demonstrate robust global exponential stability of the ET-MHE scheme, assuming a suitable detectability condition is met. In addition, we show that with the adoption of a varying horizon length, a tighter bound on the estimation error can be achieved.
Finally, we validate the effectiveness of the proposed method through two illustrative examples.
\end{abstract}


\maketitle

\section{INTRODUCTION}

\label{sec:introduction}
Moving horizon estimation (MHE) is a state estimation method that works
by minimizing a cost function that considers past measurements. This method has proven to be a powerful solution to the state estimation problem due to its applicability to general nonlinear and potentially
constrained systems subject to model inaccuracies and measurement noise.  Furthermore, strong theoretical guarantees such as robust stability properties have been shown under a mild detectability condition
(incremental input/output-to-state stability (i-IOSS)), cf. \cite{All21,Ji16,Knu23,Mul17,Sch23}. 

In certain applications, limited  resources such as computation power, energy, and communication bandwidth present significant challenges. For example, networked control systems that are  composed of numerous interconnected devices (e.g., sensors and actuators), often face bandwidth constraints that hinder simultaneous data transmission \cite{Zou21}. Moreover, in many applications it is desirable to conserve the energy of battery-operated devices \cite{Aky02}. Event-triggered methods can be employed to address these challenges by reducing the amount of unnecessary transmissions, \textcolor{rev}{see e.g. \cite{Ge20,Pen18}}.

For linear systems, significant progress has been made in event-triggered, optimization-based state estimation. An event-triggered maximum likelihood estimation  method for detectable  systems under  Gaussian noise was proposed in \cite{Shi14}. Furthermore, in works such as \cite{Yin20}, \cite{Zou20} and \cite{LiXu23} event-triggered moving horizon estimation (ET-MHE) schemes were  developed that guarantee  boundedness of the estimation error for observable systems subject to bounded disturbances.  

In the context of nonlinear systems, adaptations of the Kalman filter form the basis for several event-triggered state estimation methods. For instance, a cubature Kalman filter with a deterministic event-triggering rule was presented in \cite{Li21}. Additionally, stochastic event-triggering mechanisms in unscented and cubature Kalman filters are proposed  in \cite{Li19} and \cite{Li23}, where stochastic stability of the estimator is shown under certain observability assumptions. 
\textcolor{rev}{Other event-triggered state estimation methods for specific classes of nonlinear systems have also been proposed in the literature. For instance, state-affine systems are considered in \cite{Son21}, while an event-triggered impulse observer for nonlinear Lipschitz systems is presented in \cite{Eti16}.   }
For nonlinear systems, only a few results exist on event-triggered moving horizon estimators. In \cite{Gao22}, a centralized ET-MHE scheme was proposed for nonlinear dynamical  networks with sector-bounded nonlinearities and linear output functions. 
In \cite{Jin14, Zha14}, event-triggered MHE was studied for distributed nonlinear systems, however the event-triggering mechanism governs the communication between subsystems and does not address event-triggered transmission between a plant and a remote estimator.
In our previous work \cite{Kra24}, a robustly stable event-triggered moving horizon estimation (ET-MHE) scheme was introduced for remote state estimation of general nonlinear detectable systems. In that work, the optimization problem is required to be explicitly solved only when an event occurs. This approach not only  reduces the computational complexity of the method but also reduces the frequency with which the communication channel between the plant and a remote estimator is accessed. 
However, the method in \cite{Kra24} requires to transmit a (potentially long) sequence of measurements to the remote estimator when an event is triggered. 
This may be problematic under limited communication bandwidth \cite{Bat09}, and furthermore increases the risk of unsuccessful transmission, as transmission failure probability typically grows with packet length \cite{Que10}\color{black}. 
\par 
\color{resl} Therefore, we propose the first ET-MHE scheme for remote state estimation of general nonlinear systems that requires transmitting only the most recent measurement instead of a sequence of measurements. Hence, it not only reduces how often the communication channel needs to be accessed, but also, unlike \cite{Kra24},  can potentially largely reduce the amount of data that needs to be sent to the estimator and avoids sending large data packages. 
\color{black} 
For the proposed method we establish robust global exponential stability (RGES) of the estimation error with respect to disturbances and noise. Furthermore, we show how to modify the proposed ET-MHE by considering a varying horizon length, such that our theoretical analysis leads to a tighter bound on the estimation error. 
\par The rest of the paper is organized as follows. In Section~\ref{sec:Setup} we explain the setting of this work and provide some technical definitions. In Section~\ref{sec:MHEScheme} the ET-MHE scheme is presented. We formulate the  optimization problem and propose the event-triggering mechanism (ETM) used in the algorithm. This is followed by the stability analysis of the proposed scheme in Section \ref{sec:StabAna}. There, we show robust stability of the proposed ET-MHE scheme under a suitable detectability assumption. In Section~\ref{sec:varN} we extend the ET-MHE scheme proposed in Section~\ref{sec:MHEScheme}  such that a varying horizon length is used, resulting in tighter error bounds. Here, we again establish RGES of the estimator. Finally, Section  \ref{sec:sim} presents simulation examples to illustrate the effectiveness of the proposed method.

\section{Preliminaries and Problem Setup}
\label{sec:Setup}
We denote the set of all nonnegative real numbers by  $\mathbb{R}_{\geq0}$. The set of integers within the interval  $[a,b]$  for some  \color{resl} integers $a,b$ \color{black} is represented by  $\mathbb{I}_{[a,b]}$, while $\mathbb{I}_{\geq a}$ denotes the set of integers greater than or equal to $a$  for some  $a\in \mathbb{R}$.  
The bold symbol $\mathbf{u}$ refers to a sequence of the vector-valued variable  \mbox{${u\in\mathbb{R}^m}, \ \mathbf{u}= \{u_0,u_1,\ldots\}$} and the notation $(\mathbb{R}^m)^\infty$ denotes the set of all
sequences $\mathbf{u}$ with infinite length.
The Euclidean norm of a vector $x \in \mathbb{R}^n$  is denoted by $||x||$.  Additionally, we denote $||x||_P^2=x^\top Px$  where $P$ is a symmetric and positive definite matrix. 
The symbols $\lambda_{\text{min}}(P)$ and  $\lambda_{\text{max}}(P)$ represent the minimum and maximum eigenvalues of the matrix $P$, respectively, while $\lambda_{\text{max}}(P,Q)$ denotes the maximum generalized eigenvalue for the positive definite matrices $P$ and $Q$. Lastly, $P\succ 0$ and $P\succeq 0$ indicate that a matrix $P$ is positive definite and positive semi-definite, respectively.
\par We consider the discrete-time nonlinear system 
\begin{align}
	\begin{aligned}
		x_{t+1}&=f(x_t,u_t,w_t) \\
		y_t&=h(x_t,u_t,w_t)\\
	\end{aligned}
	\label{eq:sys}
\end{align}
with state $x_t \in \mathbb{X} \subseteq \mathbb{R}^n$, control input $u_t\in \mathbb{U} \subseteq \mathbb{R}^m$, process disturbance and measurement noise\footnote{\textcolor{rev}{Here we use $w$ to denote both process and measurement noise. This is a more \emph{general} formulation of the problem that includes, e.g., having separate additive disturbances as a special case.}}  $w_t \in \mathbb{W} \subseteq \mathbb{R}^q$ with \mbox{$0 \in \mathbb{W}$}, noisy output measurement $y_t \in \mathbb{Y} \subseteq \mathbb{R}^p$, time $t \in \mathbb{I}_{\geq 0}$, and nonlinear continuous functions $f: \mathbb{X} \times \mathbb{U} \times \mathbb{W} \rightarrow \mathbb{X}, \ h: \mathbb{X} \times \mathbb{U} \times \mathbb{W}\rightarrow \mathbb{Y}$ representing the system dynamics and the output model, respectively\footnote{\textcolor{rev}{The sets $\mathbb{X},\mathbb{W},\mathbb{U},\mathbb{Y}$ are inherently satisfied due to the physical nature of the system such as non-negativity of the absolute temperature, partial pressures, or concentrations of species in a chemical reaction, compare e.g. the discussion in \cite[Section 3]{All21}.}}. \color{fin} The input $u_t$ is considered a known input and is assumed to be available both to the plant side and to the remote estimator. \color{black}
\par To design a robustly stable MHE, it is necessary to make a suitable detectability assumption about the system. As we aim for robust global exponential stability, we use exponential i-IOSS as the detectability criterion.
\begin{ass}[Exponential i-IOSS]
	\label{ass:eIOSSET}
	The system (\ref{eq:sys})  is exponentially i-IOSS, i.e., there exist $P_1,P_2 \succ 0$, $Q, R \succeq 0$ and $\eta \in [0, 1)$ such that for any pair of initial conditions $x_{0}$, $\tilde{x}_{0}\in \mathbb{X}$ and   any input trajectories  $\mathbf{w},\mathbf{\tilde{w}}\in\mathbb{W}^\infty$ and  $\mathbf{u}\in\mathbb{U}^\infty$  it holds for all $t\geq 0$
	\begin{align}
		\begin{aligned}
			&||x_t-\tilde{x}_t||_{P_1}^2\leq ||x_0-\tilde{x}_0||_{P_2}^2 \eta^t \\&+\sum_{j=0}^{t-1}\eta^{t-j-1}||w_j-\tilde{w}_j||_Q^2
			+\sum_{j=0}^{t-1}\eta^{t-j-1}||y_j-\tilde{y}_j||_R^2,
		\end{aligned}
		\label{eq:ioss}
	\end{align}
	where $x_{t+1}=f(x_{t},u_{t},w_{t})$, $\tilde{x}_{t+1}=f(\tilde{x}_{t},u_{t},\tilde{w}_{t})$ and $y_t=h(x_{t},u_{t},w_{t})$, $\tilde{y}_t=h(\tilde{x}_{t},u_{t},\tilde{w}_{t})$.
\end{ass}
\par Note that exponential i-IOSS 
is a necessary condition for the existence of robustly exponentially stable state estimators (cf. \cite[Prop. 3]{Knu20}, \cite[Prop. 2.6]{All20}).
Adapting \cite[Theorem 8]{All20} shows that Assumption~\ref{ass:eIOSSET}  is equivalent to the existence of a quadratically-bounded i-IOSS Lyapunov function $W_{\delta}(x,\tilde{x})$ satisfying for all $x,\tilde{x} \in \mathbb{X}$, all $u\in \mathbb{U}$, all $w,\tilde{w} \in \mathbb{W}$ and all $y,\tilde{y} \in \mathbb{Y}$ 
\begin{align}
	\begin{aligned}
		&	||x-\tilde{x}||_{P_1}^2\leq W_{\delta}(x,\tilde{x})\leq	||x-\tilde{x}||_{P_2}^2,\\
		&	W_{\delta}(f(x,u,w),f(\tilde{x},u,\tilde{w}))\\ & \quad \leq \eta W_{\delta}(x,\tilde{x})+ 	||w-\tilde{w}||_{Q}^2+	||y-\tilde{y}||_{R}^2 
	\end{aligned}
	\label{eq:Lyap}
\end{align}	
with $\eta \in [0, 1)$, $P_1,P_2 \succ 0$ and $Q,R \succeq 0$.
Systematic approaches for calculating an i-IOSS Lyapunov function using LMI conditions are presented in \cite[Section IV]{Sch23,Are23}\color{black}, thereby allowing for the systematic verification of Assumption~\ref{ass:eIOSSET}.
\par In Section \ref{sec:StabAna}, we show  that the proposed ET-MHE is robustly globally exponentially stable according to the following definition.
\begin{defi}[RGES {\cite[Def. 1]{Knu18}}]
	\label{def:rges}
	A state estimator for system (\ref{eq:sys}) is robustly globally exponentially stable (RGES) if there exist $C_x,C_w>0$ and $\lambda_x,\lambda_w \in [0,1)$  such that for any initial conditions $x_0,\hat{x}_0 \in \mathbb{X}$ and any disturbance sequence~$\mathbf{w}\in\mathbb{W}^\infty$ the resulting state estimate $\hat{x}_t$ satisfies the following  for all $t\geq 0$
	\begin{align}
		\begin{aligned}
			||x_t-\hat{x}_t||\leq C_x||x_0-\hat{x}_0||\lambda_x^t+\sum_{j=0}^{t-1}
			C_w||w_j||\lambda_w^{t-j-1}.
		\end{aligned}
		\label{eq:rges}
	\end{align} 
\end{defi}
In this definition of RGES the influence of past disturbances on the error bound is discounted. As a result, the estimation error converges to zero when disturbances fade. 

\section{ET-MHE Scheme}
\label{sec:MHEScheme}
MHE is an optimization-based state estimation method that, at each time $t$, computes the current state estimate by solving an optimization problem defined over a window of past inputs and outputs.
In this section, we propose an ET-MHE scheme in which, when the optimization problem is solved, it uses a fixed window length. 
An extension of this scheme, where the horizon length is allowed to vary, is presented in a subsequent section.
\par The ET-MHE scheme presented in this section works as follows. When there is an event, a single measurement sample is transmitted from the plant side to the remote estimator. The estimator then -- and only then -- solves the nonlinear optimization problem that yields the optimal state estimate. We show that, when the optimization problem is not solved, an open-loop prediction provides the optimal state estimate.
\par The ETM is responsible for determining if an event is scheduled by computing the value of the binary event-triggering variable $\gamma_{t}$. If $\gamma_{t}=1$, an event occurs and the output measurement\footnote{\textcolor{rev}{Note that we send $y_{t-1}$ when $\gamma_t=1$ since we use the prediction form of MHE and thus optimize over the window $[t-M_t,t-1]$ to obtain the current  state estimate at time $t$ (see e.g. \cite[Chapter 3]{All18}).}} $y_{t-1}$ is sent to the estimator; $\gamma_t=0$ indicates that there is no event and thus no data transmission at time $t$.
The most recent  time an event was scheduled before the current time $t$ is denoted by 
$\epsilon_t\coloneqq\max\{0\leq\tau<t|\gamma_{\tau}=1\}$. For simplicity, we set $\epsilon_0=0$ and $\gamma_0=1$. Furthermore, $\delta_t$  refers to the number of time steps  that have passed since the last event, i.e.,
\begin{align*}
	\delta_{t}=	\begin{cases}
		t-\epsilon_t, & \gamma_t=0\\
		0, &\gamma_t=1.
	\end{cases}
\end{align*}
\color{resl} These event-related variables are summarized in Table~\ref{tab:var} at the end of this section. \color{black}
How the ETM schedules an event will be discussed in more detail below.
\par As remote estimator we consider an MHE with a window length  $M_t=\min\{t,M+\delta_t\}$ with $M\in \mathbb{I}_{\geq 0}$.
\begin{rem}
	Note that within our scheme, the MHE optimization problem is required to be explicitly solved only when an event is triggered (see Proposition~\ref{prop:NLPOL} below).  At those time instances $\delta_t=0$, which implies that the horizon length is equal to $M$ for $t\geq M$. Therefore, whenever the optimization problem is explicitly solved, it is exclusively solved with a fixed horizon length of $M$ for $t\geq M$.
	\label{rem:M}
\end{rem}
Notice that we still define a variable horizon $M_t$ for all $t$ because it simplifies our analysis of the proposed ET-MHE.  \color{resl} We first show that (i) solving the optimization problem at every time instance with variable horizon length, and (ii)  solving it only at time instances where an event is triggered, and using an open-loop prediction between two event times, are in fact equivalent approaches. \color{black}
%
%
Then,  we simply show robust stability of the former case (see Theorem~\ref{thm:stab2}).
\par
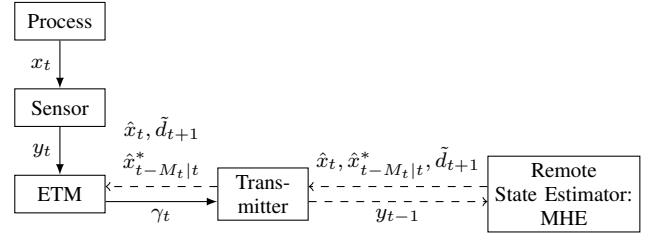
\begin{figure}[!t]
\vspace{0.55em}
\centering
\centerline{\begin{tikzpicture}
	\usetikzlibrary{arrows}
	\usetikzlibrary{shapes.geometric}
\begin{footnotesize}
	\tikzstyle{block} = [draw, minimum width=1.2cm, minimum height=0.5cm,  
	align = center]
	[shorten -2pt,shorten -2pt,-implies]
	\node[block] (sys)  at (-3,3.3) {Process};
	\node[block] (sens)  at (-3,2.15) {Sensor};
	\node[block] (ES) at (-3,1) {ETM};
	\node[block] (TM) at (-0.3,1) {Trans-\\mitter};
	\node[block] (RO)  at (3.7,1) {Remote\\ State Estimator: \\ MHE};
	\draw[-latex] (sys) --(sens) node [left,midway] {$x_{t}$ };
	\draw[-latex] (sens) --(ES) node [left,midway] { $y_{t}$ };
	\draw[-latex] ([yshift=-1mm]ES.east) --([yshift=-1mm]TM.west) node [below,midway] {$\gamma_{t}$};
	\draw[dashed, ->] ([yshift=1mm]TM.west) --([yshift=1mm]ES.east) node [above,midway] {$\begin{aligned}\hat{x}_{t}, \tilde{d}_{t+1}\\ \hat{x}^*_{t-M_t|t}\end{aligned}$ };
	\draw[dashed, ->] ([yshift=-1mm]TM.east) --([yshift=-1mm]RO.west) node [below,midway] {$y_{t-1}$};
	\draw[dashed, ->] ([yshift=1mm]RO.west) --([yshift=1mm]TM.east) node [above,midway] {$\hat{x}_{t},\hat{x}^*_{t-M_t|t},\tilde{d}_{t+1}$};
\end{footnotesize}
\end{tikzpicture}}
\caption{Diagram of the overall ET-MHE framework. The dashed arrows represent the communication that is only required in case of an event.} 
\label{fig:system2}
\end{figure}

 Since in this scheme we only send a single measurement in case of an event, the MHE cannot access all the measurements in the estimation horizon. Therefore, we define the set $K_s$ that contains the time instances of the measurements that were sent to the remote estimator, i.e., $K_s\coloneqq\{\tau|\gamma_{\tau+1}=1\}$. 
The overall ET-MHE framework is depicted in Figure~\ref{fig:system2}.
\par The MHE's nonlinear program (NLP) for $t<\tilde{\tau}$ with $\tilde{\tau}\coloneqq\min\{\tau|\gamma_{\tau}=0\}$ is given by 
\begin{subequations}
	\label{eq:ET2ETNLP2}
	\begin{align}
		\min_{\hat{x}_{t-M_t|t}, \hat{w}_{\cdot|t}} 
		&	J(\hat{x}_{t-M_t|t}, \hat{w}_{\cdot|t},\hat{y}_{\cdot|t},t)\\
		\text{s.t.} \  \hat{x}_{j+1|t}&=f(\hat{x}_{j|t},u_j,\hat{w}_{j|t}), \ j\in \mathbb{I}_{[t-M_t,t-1]},\label{eq:NLPC1}\\
		\hat{y}_{j |t}&=h(\hat{x}_{j|t},u_j,\hat{w}_{j|t}), \ j\in \mathbb{I}_{[t-M_t,t-1]}
		,\label{eq:NLPC2}\\
		\hat{w}_{j|t} &\in \mathbb{W}, \hat{y}_{j|t} \in \mathbb{Y}, \ j\in \mathbb{I}_{[t-M_t,t-1]},\label{eq:NLPC3}\\
		\hat{x}_{j|t} &\in \mathbb{X}, \ j\in \mathbb{I}_{[t-M_t,t]} \label{eq:NLPCe}
	\end{align}
\end{subequations}
The estimated state for time $j$ 
computed at the current time $t$  is indicated by the notation $\hat{x}_{j|t}$. Analogously,  $\hat{w}_{j|t}$ and $\hat{y}_{j|t}$ denote the  estimated disturbances and outputs.
The notations  $\hat{x}^*_{\cdot|t}$, $\hat{w}^*_{\cdot|t}$ and  $\hat{y}^*_{\cdot|t}$ refer to
the optimal state, disturbance, and output sequences, respectively,  that minimize the cost function $J$.
The optimal estimate at the current time $t$ is defined as $\hat{x}_t \coloneqq \hat{x}^*_{t|t}$. Moreover, the notation $\hat{e}_t\coloneqq x_t-\hat{x}_t$ is used to describe the estimation error at time $t$.
For $t\geq\tilde{\tau}$, the MHE optimization problem  is given by 	(\ref{eq:ET2ETNLP2}) and the following additional constraint
\begin{align}
	\begin{aligned}
		&\sum_{\substack{j\in \mathbb{I}_{[t-M_t,\mu_t-1]} \setminus K_s}}\eta^{\mu_t-j-1} ||\hat{y}_{j |t}-\tilde{y}_{j |\mu_t-1}||_R^2\\\leq& \alpha \Big( \sum_{j=\mu_t-M_{\mu_{t}}}^{\epsilon_{\mu_t}-1} \eta^{\mu_t-j-1}2 ||\hat{w}_{j|\epsilon_{\mu_t}}^*||_Q^2 \\+& \sum_{\substack{j\in \mathbb{I}_{[\mu_t-M_{\mu_{t}},\epsilon_{\mu_t}-1]} \cap K_s}}\eta^{\mu_t-j-1}||\hat{y}_{j|\epsilon_{\mu_t}}^*-y_j||_R^2\Big) 
	\end{aligned}
	\label{eq:NLPC4}
\end{align}
with \begin{align*}
	\tilde{y}_{j |\mu_t-1}=\begin{cases}
		\hat{y}^*_{j |\mu_t-1}, &j<\mu_t-1\\
		h(\hat{x}_{\mu_t-1},u_{\mu_t-1},0), &j=\mu_t-1.	
	\end{cases}
\end{align*}
The parameters $\alpha,\eta, Q,R$ correspond to the parameterization of the cost function (see below).
The time instant $\mu_t$, defined for $t\geq\tilde{\tau}$, refers to the last time  when no event was triggered, i.e., $\mu_t\coloneqq\max\{\tau | \tilde{\tau} \leq \tau \leq t, \gamma_\tau = 0\}$. 
Note also that $\mu_t-M_{\mu_{t}}=\epsilon_{\mu_t}-M_{\epsilon_{\mu_t}}$ since  $\delta_{\epsilon_{\mu_t}}=0$.  
\color{resl}
The additional constraint (\ref{eq:NLPC4}) is only required for $t\geq \tilde{\tau}$, since for $t<\tilde{\tau}$, by definition of $\tilde{\tau}$, an event was triggered at every time step and thus the output $y_t$ was transmitted at every time step. \color{black}
 Consequently,  the ET-MHE scheme corresponds to the non event-triggered MHE case for all $t<\tilde{\tau}$.
This constraint can be interpreted as follows. The MHE optimization problem should drive the (complete) estimated output sequence towards meaningful values. At the times when there are measurements available, the estimated outputs should approach the measurements (this requirement is given by the cost function (\ref{eq:objfunc2}) below). At the times when there are no measurements available, the estimated outputs should be close to a previously obtained optimal output sequence. This is achieved by the constraint (\ref{eq:NLPC4}), which forces the left-hand side to be small depending on how accurate $\hat y^*_{j|\epsilon_{\mu_t}}$ was.
\color{new} As it will be shown in Section~\ref{sec:StabAna}, the additional constraint (\ref{eq:NLPC4}) is used in the stability analysis to facilitate the derivation of  finite-horizon error bounds for the case of  an event being triggered,  and thus  enables us to prove robust stability for the proposed scheme. 
\color{resl}
Moreover, after introducing the ETM later in this section, we show (see Lemma~\ref{lem:addC}) that under the proposed triggering condition the true system trajectories necessarily satisfy the additional constraint (\ref{eq:NLPC4}). \color{black}
This is an important characteristic that justifies to require (\ref{eq:NLPC4}) also for the optimal solution of the optimization problem. In fact, in numerical experiments (see Section~\ref{sec:sim}) we observed that this constraint  did not affect the performance of the estimator in a significant way.
\color{black}
\par We consider the following cost function
\begin{align}
\begin{aligned}
	J&(\hat{x}_{t-M_t|t}, \hat{w}_{\cdot|t},  \hat{y}_{\cdot|t},t)=2\eta^{M_t}||\hat{x}_{t-M_t|t}-\hat{x}_{t-M_t}||_{P_2}^2 \\&+\max\{1,\alpha\}\Big(\sum_{j=t-M_t}^{t-1} \eta^{t-j-1}2 ||\hat{w}_{j|t}||_Q^2\\& +\sum_{j\in \mathbb{I}_{[t-M_t,t-1]}\cap K_s} \eta^{t-j-1} ||\hat{y}_{j|t}-y_j||_R^2 \Big).
\end{aligned}
\label{eq:objfunc2}
\end{align}
The parameter $\alpha\geq0$ is a design variable and corresponds to the sensitivity of the event-triggering condition, see Remark~\ref{rem:alpha} below.
The cost function consists of two parts, namely the prior weighting and the stage cost. The prior weighting  penalizes the difference between the first element of the estimated state sequence  $\hat{x}_{t-M_t|t}$ and the prior estimate $\hat{x}_{t-M_t}$ that was obtained  at time  $t-M_t$. 
The estimated noise and the difference between the measured and the estimated output are penalized in the  stage cost \color{resl} with the weighting matrices $Q$ and $R$, respectively. 
The parameters $P_2,Q,R$ and $\eta$ correspond to the ones  in Assumption~\ref{ass:eIOSSET}. 
This is leveraged to ensure RGES in Theorem~\ref{thm:stab2} below. However, assuming a  parameterization of the cost function consistent with the detectability condition of exponential i-IOSS does not restrict any tuning possibilities. \color{black} If the system is  exponentially i-IOSS, then the cost function can be parameterized using any  positive definite matrices $P_2,Q$ and $R$ (since (\ref{eq:Lyap}) can be rescaled accordingly, cf. \cite[Remark 1]{Sch23}).
The discount factor $\eta$ reduces the influence of disturbances and output measurements further in the past. A discounting MHE cost function was introduced in \cite{Knu18}, and has proven particularly useful in the robust stability analysis of MHE.
\par  
The following proposition from \cite{Kra24} 
shows why it is sufficient to only solve the optimization problem at the time instances when an event occurs. When there is no event, the current estimate can be obtained using open-loop predictions.
The proof requires only minor modifications to account for the fact that only  measurements at time instances 
$t\in K_s$ are transmitted to the remote side. Nevertheless, we include it here for the sake of completeness.
\begin{prop}[\!{\cite[Proposition 1]{Kra24}}]%
	\label{prop:NLPOL}	
The solution of the NLP (\ref{eq:ET2ETNLP2})-(\ref{eq:NLPC4}) at time $t\geq0$ is given by 
\begin{subequations}
	\label{eq:Prop}
	\begin{align}
		\hat{x}^*_{t-M_t|t} &= \hat{x}^*_{t-\delta_t-M_{t-\delta_t}|t-\delta_t},\\
		\hat{w}^*_{j|t} &= \hat{w}^*_{j|t-\delta_t}, \ j\in \mathbb{I}_{[t-M_t,t-\delta_t-1]},\\
		\hat{w}^*_{j|t} &= 0, \ j \in  \mathbb{I}_{[t-\delta_t,t-1]}. \label{eq:Propw0}
	\end{align}
\end{subequations}
\end{prop}
\begin{proof}
First,	note that the last term of the cost function (\ref{eq:objfunc2}) only considers the time instances when the corresponding measurements were transmitted to the remote estimator, i.e, the measurements in $[t-M_t,t-\delta_t-1]\cap K_s$. Thus,   $\hat{w}_{j|t}=0$ for all \mbox{$j\in [t-\delta_t,t-1]$} minimizes $J(\hat{x}_{t-M_t|t}, \hat{w}_{\cdot|t},  \hat{y}_{\cdot|t},t)$. 
Hence, the following holds for the optimal value of the cost function at time $t$ 
\begin{align*}
	\begin{aligned}
		&J(\hat{x}_{t-M_t|t}^*, \hat{w}_{\cdot|t}^*,  \hat{y}_{\cdot|t}^*,t)\\=&\eta^{\delta_t} J(\hat{x}_{t-\delta_t-M_{t-\delta_t}|t-\delta_t}^*, \hat{w}_{\cdot|t-\delta_t}^*,  \hat{y}_{\cdot|t-\delta_t}^*,t-\delta_t)
	\end{aligned}
\end{align*}  
where  $t-\delta_t-M_{t-\delta_t}=t-M_t$ since $\delta_{t-\delta_t} =0$. Thus, the solution obtained by  minimizing  $J(\hat{x}_{t-M_t|t}, \hat{w}_{\cdot|t},\hat{y}_{\cdot|t},t)$ is given by (\ref{eq:Prop}). 
\end{proof}
\par According to Proposition~\ref{prop:NLPOL}, if $\gamma_t=0$, i.e., no event occurs, 
the estimates $\hat{x}_j$  for all $j\in [t-\delta_t+1,t]$ are given by an open-loop prediction
\begin{align}
\hat{x}_{j}=f(\hat{x}_{j-1},u_{j-1},0).
\label{eq:OL}
\end{align}
Consequently, the NLP~(\ref{eq:ET2ETNLP2}) only needs to be explicitly solved when an event is triggered. Since $\gamma_t=1$ implies $\delta_t=0$, the NLP that must be solved explicitly has a fixed horizon length $M$ for $t\geq M$, as discussed in  Remark \ref{rem:M}.
\par \textcolor{rev}{As we show later in the proof of Theorem~\ref{thm:stab2}, RGES can be established using the following ETM, that determines the value of the scheduling variable $\gamma_t$ at each time}
\begin{align}
\gamma_{t}=	\begin{cases}
	0, & \text{if} \ \quad  \begin{aligned}&\sum_{j\in \mathbb{I}_{[\epsilon_t-M_{\epsilon_t},{\epsilon_t-1}]}\setminus K_s}\eta^{t-j-1} ||y_j-\hat{y}_{j|\epsilon_t}^*||_R^2\\&+\sum_{j=\epsilon_t}^{t-1}\eta^{t-j-1} ||y_j-h(\hat{x}_j,u_j,0)||_R^2 \\&< \alpha \eta^{t-\epsilon_t}d_t
	\end{aligned}\\
	1, &\text{otherwise}
\end{cases}
\label{eq:ET2trigruled2}
\end{align}
 and
\begin{align}
\begin{aligned}
	d_{t}&=	2\sum_{j=\epsilon_t-M_{\epsilon_t}}^{\epsilon_t-1} \eta^{\epsilon_t-1-j} ||\hat{w}_{j|\epsilon_t}^*||_Q^2\\&+ \sum_{j\in \mathbb{I}_{[\epsilon_t-M_{\epsilon_t},{\epsilon_t-1}]}\cap K_s} \eta^{\epsilon_t-1-j}||y_j-\hat{y}_{j|\epsilon_t}^*||_R^2.
\end{aligned}
\label{eq:dwy}
\end{align}
However, directly evaluating the condition above would require sending several estimated outputs back to the plant side.
In many practical scenarios, receiving data consumes less  energy than transmitting at a wireless sensor node \cite{Che17,Al24}.
Nevertheless, we still want to reduce the amount of data that needs to be sent back as much as possible.  \color{black} 
Hence, we modify the ETM (\ref{eq:ET2trigruled2}) by exploiting the following bound
\begin{align}
\begin{aligned}
	&\sum_{j\in \mathbb{I}_{[\epsilon_t-M_{\epsilon_t},{\epsilon_t-1}]}\setminus K_s}\eta^{t-j-1} ||y_j-\hat{y}_{j|\epsilon_t}^*||_R^2
	\\ \leq& 2	\sum_{j\in \mathbb{I}_{[\epsilon_t-M_{\epsilon_t},{\epsilon_t-1}]}\setminus K_s}\eta^{t-j-1} ||\bar{y}_{j}^{\epsilon_t}-\hat{y}_{j|\epsilon_t}^*||_R^2\\&+2	\sum_{j\in \mathbb{I}_{[\epsilon_t-M_{\epsilon_t},{\epsilon_t-1}]}\setminus K_s}\eta^{t-j-1} ||y_j-\bar{y}_{j}^{\epsilon_t}||_R^2\\
\end{aligned}
\label{eq:boundtrigcond}
\end{align}
\vspace{1em}
with  $\bar{x}_{\epsilon_t-M_{\epsilon_t}}^{\epsilon_t}=\hat{x}_{\epsilon_t-M_{\epsilon_t}|\epsilon_t}$ and
\begin{align*}
\begin{aligned}
	&\bar{x}_{j+1}^{\epsilon_t}=f(\bar{x}_j^{\epsilon_t},u_j,0),\\&\bar{y}_j^{\epsilon_t}=h(\bar{x}_j^{\epsilon_t},u_j,0), \quad j\in[\epsilon_t-M_{\epsilon_t},\epsilon_t-1].
\end{aligned}
\end{align*} 
We define \begin{align}
p_t=\sum_{j\in \mathbb{I}_{[\epsilon_t-M_{\epsilon_t},{\epsilon_t-1}]}\setminus K_s}\eta^{\epsilon_t-j-1} ||\bar{y}_{j}^{\epsilon_t}-\hat{y}_{j|\epsilon_t}^*||_R^2.
\label{eq:pt}
\end{align}
Note that both $d_t$ and $p_t$ as defined in (\ref{eq:dwy}) and (\ref{eq:pt}), respectively, can be calculated at the remote side since all required quantities to this end have been sent to the remote side.
By sending additionally the first element of the estimated state sequence back to the plant side, $\bar{y}_{j}^{\epsilon_t}$ can be calculated at the plant side. 
Using (\ref{eq:boundtrigcond}) we can now formulate our event-triggering condition as follows
\begin{align}
\gamma_{t}=	\begin{cases}
	0, & \text{if} \ \quad  \begin{aligned}&2	\sum_{j\in \mathbb{I}_{[\epsilon_t-M_{\epsilon_t},{\epsilon_t-1}]}\setminus K_s}\eta^{t-j-1} ||y_j-\bar{y}_{j}^{\epsilon_t}||_R^2\\&+\sum_{j=\epsilon_t}^{t-1}\eta^{t-j-1} ||y_j-h(\hat{x}_j,u_j,0)||_R^2 \\&< \eta^{t-\epsilon_t}\tilde{d}_t
	\end{aligned}\\
	1, &\text{otherwise}
\end{cases}
\label{eq:ET2trigruledAdap}
\end{align}
where $\tilde{d}_t=(\alpha d_t-2p_t)$.
As previously stated, if $\gamma_{t}=1$, the measurement $y_{t-1}$ is transmitted to the remote state estimator and the optimization problem is solved at time $t$. 
Then, both $d_{t+1}$ and $p_{t+1}$ can be computed using the estimated disturbance and output sequences $\hat{w}_{\cdot|t}^*, \hat{y}_{\cdot|t}^*$ since $\epsilon_{t+1}=t$.  Thereafter, the scalar $\tilde{d}_{t+1}$ together with the first and the last element of the estimated state sequence, i.e., $\hat{x}^*_{t-M_t|t}$ and $\hat{x}_t$ is sent back to the ETM to evaluate the triggering condition for the next time step (cf. Figure \ref{fig:system2}).
Note that $d_t$ and $p_t$, and thus $\tilde{d}_t$, are constant between events, i.e., $d_{t+1}=d_t$ and $p_{t+1}=p_t$  if $\gamma_t=0$. This follows from (\ref{eq:Prop}) and  the fact that $\epsilon_{t+1}=\epsilon_t$ if $\gamma_t=0$. The  steps of the ET-MHE scheme are outlined in  Algorithm~\ref{alg2}. 
\floatstyle{spaceruled}
\restylefloat{algorithm}
\begin{algorithm}[!t]
\caption{Event-triggered MHE}
\begin{algorithmic}[1]
	\State Set $\gamma_0=1$ and $\tilde{d}_{1}=0$.
	\State Set $t=1$.
	\State ETM computes $\gamma_{t}$.
	\If{$\gamma_{t}=1$}
	\State $y_{t-1}$ is sent to remote estimator.
	\State NLP (\ref{eq:ET2ETNLP2})-(\ref{eq:objfunc2}) of MHE is solved.
	\State $\tilde{d}_{t+1}$ is calculated 
	\State $\tilde{d}_{t+1}$, $\hat{x}^*_{t-M_t|t}$ and $\hat{x}_t$ are sent back to the ETM. 
	\Else 
	\State   $\hat{x}_t$ is calculated according to (\ref{eq:OL}). 
	\EndIf
	\State  $\hat{y}_t=h(\hat{x}_t,u_t,0)$ is calculated.
	\State Set $t=t+1$ and go back to step 3.
\end{algorithmic}
\label{alg2}
\end{algorithm}
Notice that in the algorithm we set $\tilde{d}_{1}=0$ and hence there is always an event triggered at time $t=1$, i.e., the first measurement $y_0$ is sent to the remote side. 
\begin{rem}
	\label{rem:alpha}
In the proposed method,  $\alpha$ is a design parameter that affects the frequency of an event being triggered. 
When  $\alpha$ is increased, the occurrence of events decreases, which consequently enlarges the disturbance gain in the error bound derived in Theorem~\ref{thm:stab2} below.
\color{resl} Hence, the tuning parameter $\alpha$ is chosen according to the desired trade-off between estimation performance and triggering rate.\color{black}
\end{rem}
Having introduced the optimization problem and the event-triggering condition, we now state the following lemma, which shows that under ETM  (\ref{eq:ET2trigruledAdap}) the true system trajectories always satisfy constraint (\ref{eq:NLPC4}).
\begin{lem}	
	\label{lem:addC}
	If the transmission instants are determined according to the ETM (\ref{eq:ET2trigruledAdap}), then the true system trajectories satisfy constraint (\ref{eq:NLPC4}).
\end{lem}
\begin{proof}
	Since $\gamma_{\mu_t}=0$, the inequality in (\ref{eq:ET2trigruledAdap}) holds at $\mu_t$. 
	This implies that the inequality in (\ref{eq:ET2trigruled2}) is satisfied as well, which follows from the developments between (\ref{eq:ET2trigruled2}) and (\ref{eq:ET2trigruledAdap}).	Hence,
	\begin{align}
		\begin{aligned}
			&\sum_{j\in \mathbb{I}_{[\epsilon_{\mu_t}-M_{\epsilon_{\mu_t}},{\epsilon_{\mu_t}-1}]}\setminus K_s}\eta^{\mu_t-j-1} ||y_j-\hat{y}_{j|\epsilon_{\mu_t}}^*||_R^2\\&+\sum_{j=\epsilon_{\mu_t}}^{\mu_t-1}\eta^{\mu_t-j-1} ||y_j-h(\hat{x}_j,u_j,0)||_R^2 \\< &\alpha \big(2\sum_{j=\epsilon_{\mu_t}-M_{\epsilon_{\mu_t}}}^{\epsilon_{\mu_t}-1} \eta^{\mu_t-1-j} ||\hat{w}_{j|\epsilon_{\mu_t}}^*||_Q^2\\&+ \sum_{j\in \mathbb{I}_{[\epsilon_{\mu_t}-M_{\epsilon_{\mu_t}},{\epsilon_{\mu_t}-1}]}\cap K_s} \eta^{{\mu_t}-1-j}||y_j-\hat{y}_{j|\epsilon_{\mu_t}}^*||_R^2\big).
		\end{aligned}
		\label{eq:ET_mu}
	\end{align}
	Now  consider  the constraint (\ref{eq:NLPC4}) at time $t$ with
	$\hat{y}(j|t)$ on the left-hand side of (\ref{eq:NLPC4}) replaced by the output $y(j)$ corresponding to the true system trajectory.
	Recall that 	$\mu_t-M_{\mu_{t}}=\epsilon_{\mu_t}-M_{\epsilon_{\mu_t}}$. Thus, 
	the  right-hand side of (\ref{eq:ET_mu}) is the same as the right-hand side of (\ref{eq:NLPC4}). Notice that by Proposition~\ref{prop:NLPOL}	$\hat{y}^*_{j|\mu_t-1}=\hat{y}^*_{j|\epsilon_{\mu_t}}, \ j\in\mathbb{I}_{[\epsilon_{\mu_t}-M_{\epsilon_{\mu_t}},\epsilon_{\mu_t}-1]}$, and $\hat{y}^*_{j|\mu_t-1}=h(\hat{x}_j,u_j,0), \ j\in\mathbb{I}_{[\epsilon_{\mu_t},\mu_t-2]}$. Thus, since  $t-M_t\geq \epsilon_{\mu_t}-M_{\epsilon_{\mu_t}}$, the left-hand side of (\ref{eq:ET_mu})  upper bounds the left-hand side of (\ref{eq:NLPC4}). Therefore, (\ref{eq:ET_mu}) implies (\ref{eq:NLPC4}) evaluated at time $t$ for the true system trajectories.
\end{proof}
As discussed before, Lemma~\ref{lem:addC} shows that the constraint  (\ref{eq:NLPC4}) is meaningful, in the sense that the optimization problem can benefit from restricting the candidate solutions to a smaller set of trajectories to which the true trajectories belong.
\color{resl}
\par Table~\ref{tab:var} provides an overview of the event-related variables. The variable $\sigma_t$ is used in the developments of Section~\ref{sec:varN}.
\begin{table}[t!]
	\color{resl}
	\centering
	\caption{Summary of event-related variables}
	\label{tab:var}
	\begin{tabular}{l p{0.72\linewidth}}
		\toprule
		\textbf{Variable} & \textbf{Description} \\
		\midrule
		$\epsilon_t$       & Most recent event time before time $t$  \\
		$\delta_t$  & Number of time steps since the last event \\
		$\mu_t$  & Most recent time at which no event was triggered   \\
		$\tilde\tau$        & First time at which no event was triggered\\
		$\sigma_t$        &  Most recent time at which $\min\{t,2M\}$ consecutive events had occurred  \\
		\bottomrule
	\end{tabular}
\end{table}
\color{black}
			\section{Stability Analysis} 
			\label{sec:StabAna}
			Having introduced the ET-MHE scheme, this section now focuses on showing  robust stability of the proposed method. 
			The shift from transmitting a sequence of measurements, as in \cite{Kra24}, to just a single measurement impacts the formulation of the stage cost. This change is the main reason why we had to \textcolor{rev}{fundamentally change} the proof strategy compared to the robust stability proof in \cite{Kra24}.
			\begin{thm}[RGES of ET-MHE]
				Let Assumption \ref{ass:eIOSSET} hold and the horizon $M\in \mathbb{I}_{\geq 0}$ be chosen such that $24\lambda_{\text{max}}(P_2,P_1)\eta^{M}<1$. Then there exists $\rho \in [0,1)$ such that the state estimation error of the ET-MHE
				scheme~(\ref{eq:ET2ETNLP2})-(\ref{eq:objfunc2}) with $M_t=\min\{t,M+\delta_t\}$ and with the event scheduling condition (\ref{eq:ET2trigruledAdap}) satisfies for all $t\geq 0$ 
				\begin{align*}
					\begin{aligned}
						||\hat{e}_t||&\leq	 \sqrt{\frac{24\lambda_{\text{max}}(P_2)}{\lambda_{\text{min}}(P_1)}}\sqrt{\rho}^{t}||\hat{e}_{0}|| \\&+\sqrt{\frac{3\max\{10\alpha+2,12\}\lambda_{\text{max}}(Q)}{\lambda_{\text{min}}(P_1)}}\sum_{j=0}^{t-1} \sqrt{\rho}^{t-j-1}  ||w_j||,
					\end{aligned}
				\end{align*}
				i.e., the ET-MHE is an RGES estimator according to Definition~\ref{def:rges}.
				\label{thm:stab2}
			\end{thm}
					\begin{proof}
						Due to the constraints (\ref{eq:NLPC1})-(\ref{eq:NLPCe}) in the NLP, at each time step $t$, the estimated trajectories satisfy (\ref{eq:sys}), $\hat{x}_{j|t}\in \mathbb{X}$ for all $j\in \mathbb{I}_{[t-M_t,t]}$ and $\hat{w}_{j|t}\in \mathbb{W},\hat{y}_{j|t}\in \mathbb{Y}$ for all $j\in \mathbb{I}_{[t-M_t,t-1]}$. Thus, we can apply (\ref{eq:ioss}) to obtain
						\begin{align*}
							\begin{aligned}
								||\hat{x}_t-x_t||_{P_1}^2 &\leq \eta^{M_t} ||\hat{x}_{t-M_t|t}^*-x_{t-M_t}||_{P_2}^2\\&+\sum_{j=t-M_t}^{t-1}\eta^{t-j-1}||\hat{w}_{j|t}^*-w_j||_Q^2\\&
								+\sum_{j=t-M_t}^{t-1}\eta^{t-j-1}||\hat{y}_{j|t}^*-y_j||_R^2. 
							\end{aligned}
						\end{align*}
						Applying the Cauchy-Schwarz inequality, and Young’s inequality, it holds that
						\begin{align*}
							||\hat{w}_{j|t}^*-w_j||_Q^2\leq 2||\hat{w}_{j|t}^*||_Q^2+ 2||w_j||_Q^2
						\end{align*}
						and therefore,
						\begin{align*}
							\begin{aligned}
								||\hat{x}_t-x_t||_{P_1}^2 &\leq 
								\eta^{M_t} ||\hat{x}_{t-M_t|t}^*-x_{t-M_t}||_{P_2}^2\\&+\sum_{j=t-M_t}^{t-1}\eta^{t-j-1} 2||w_j||_Q^2\\
								&+\sum_{j=t-M_t}^{t-1}\eta^{t-j-1}2||\hat{w}_{j|t}^*||_Q^2\\&	 +\sum_{j=t-M_t}^{t-1}\eta^{t-j-1}||\hat{y}_{j|t}^*-y_j||_R^2.
							\end{aligned}
						\end{align*}
						Due to
						\begin{align*}
							\begin{aligned}
								&||\hat{x}_{t-M_t|t}^*-x_{t-M_t}||_{P_2}^2\\=& ||\hat{x}_{t-M_t}-x_{t-M_t}+\hat{x}_{t-M_t|t}^*-\hat{x}_{t-M_t}||_{P_2}^2\\
								\leq &2||\hat{x}_{t-M_t}-x_{t-M_t}||_{P_2}^2+2||\hat{x}_{t-M_t|t}^*-\hat{x}_{t-M_t}||_{P_2}^2
							\end{aligned}
						\end{align*}
						we can write
						\begin{align}
							\begin{aligned}
								||\hat{x}_t-x_t||_{P_1}^2& \leq 
								2\eta^{M_t}||\hat{x}_{t-M_t}-x_{t-M_t}||_{P_2}^2\\&+2\eta^{M_t}||\hat{x}_{t-M_t|t}^*-\hat{x}_{t-M_t}||_{P_2}^2\\&+\sum_{j=t-M_t}^{t-1}\eta^{t-j-1} 2||w_j||_Q^2
								\\&+\sum_{j=t-M_t}^{t-1}\eta^{t-j-1}2||\hat{w}_{j|t}^*||_Q^2\\&	 +\sum_{j\in \mathbb{I}_{[t-M_t,t-1]}\cap K_s}\eta^{t-j-1}||\hat{y}_{j|t}^*-y_j||_R^2\\&+\sum_{j\in \mathbb{I}_{[t-M_t,t-1]}\setminus K_s}\eta^{t-j-1} ||\hat{y}_{j|t}^*-y_j||_R^2.
							\end{aligned}
							\label{eq:ET2eiossEstsplit2}
						\end{align} 
						\textcolor{rev}{
							In the following, we aim to establish an upper bound for the last term in (\ref{eq:ET2eiossEstsplit2}). If $t< \tilde\tau$, then the last term in (\ref{eq:ET2eiossEstsplit2}) is an empty sum and thus zero. Therefore, it is sufficient to focus only on  the case $t\geq\tilde\tau$. 
							If $t\geq\tilde\tau$,} then 
						since $\mu_t$ refers to the last time no event was scheduled, it follows that the set $\mathbb{I}_{[t-M_t,t-1]}\setminus K_s$ is equal to the set $\mathbb{I}_{[t-M_t,\mu_t-1]}\setminus K_s$.
						Now, we can bound the last term of (\ref{eq:ET2eiossEstsplit2}) as follows 
						\begin{align}
							\begin{aligned}
								&\sum_{j\in\mathbb{I}_{[t-M_t,\mu_t-1]}\setminus K_s}\eta^{t-j-1}||y_j-\hat{y}_{j|t}^*||_R^2\\=&\sum_{j\in\mathbb{I}_{[t-M_t,\mu_t-1]}\setminus K_s}\eta^{t-j-1}||y_j-\hat{y}_{j|\mu_t}^*+\hat{y}_{j|\mu_t}^*-\hat{y}_{j|t}^*||_R^2\\\leq& 2\sum_{j\in\mathbb{I}_{[t-M_t,\mu_t-1]}\setminus K_s}\eta^{t-j-1} ||y_j-\hat{y}_{j|\mu_t}^*||_R^2\\+&2\sum_{j\in\mathbb{I}_{[t-M_t,\mu_t-1]}\setminus K_s}\eta^{t-j-1} ||\hat{y}_{j|\mu_t}^*-\hat{y}_{j|t}^*||_R^2
							\end{aligned}
							\label{eq:boundDeltaY}
						\end{align}
						From Proposition~\ref{prop:NLPOL} and the facts that $\mu_t-\delta_{\mu_t} = \epsilon_{\mu_t} $ and $\epsilon_{\mu_t}-M_{\epsilon_{\mu_t}} \leq t-M_t$, we get
						\begin{align*}
							\begin{aligned}
								&2\eta^{t-\mu_t}\sum_{j\in\mathbb{I}_{[t-M_t,\mu_t-1]}\setminus K_s}\eta^{\mu_t-j-1} ||y_j-\hat{y}_{j|\mu_t}^*||_R^2 \\
								=&2\eta^{t-\mu_t}\Big(\sum_{j\in \mathbb{I}_{[t-M_t,\epsilon_{\mu_t}-1]}\setminus K_s}\eta^{\mu_t-j-1} ||y_{j}-\hat{y}^*_{j |\epsilon_{\mu_t}}||_R^2\\+& \sum_{\substack{j=\epsilon_{\mu_t}}}^{\mu_t-1} \eta^{\mu_t-j-1}||y_j-h(\hat{x}_j,u_j,0)||_R^2\Big)\\\leq&
								2\eta^{t-\mu_t}\Big(\sum_{j\in \mathbb{I}_{[\epsilon_{\mu_t}-M_{\epsilon_{\mu_t}},\epsilon_{\mu_t}-1]}\setminus K_s}\eta^{\mu_t-j-1} ||y_j-\hat{y}^*_{j |\epsilon_{\mu_t}}||_R^2\\+& \sum_{j=\epsilon_{\mu_t}}^{\mu_t-1}\eta^{\mu_t-j-1}||y_j-h(\hat{x}_j,u_j,0)||_R^2\Big).
							\end{aligned}	
						\end{align*}
						If no event is triggered and thus  the bound in (\ref{eq:ET2trigruledAdap}) holds, then, by (\ref{eq:boundtrigcond}), the bound in (\ref{eq:ET2trigruled2}) holds as well.
						Therefore,  since no event is triggered at time $\mu_t$, 
						we obtain
						\begin{align}
							\begin{aligned}
								&2\eta^{t-\mu_t}\sum_{j\in\mathbb{I}_{[t-M_t,\mu_t-1]}\setminus K_s}\eta^{\mu_t-j-1} ||y_j-\hat{y}_{j|\mu_t}^*||_R^2 \\&\leq
								2\eta^{t-\mu_t}\big(\alpha \sum_{j=\mu_t-M_{\mu_t}}^{\epsilon_{\mu_t}-1}\eta^{\mu_t-j-1} 2||\hat{w}_{j|\epsilon_{\mu_t}}^*||_Q^2\\&+ \alpha
								\sum_{j\in \mathbb{I}_{[{\mu_t}-M_{{\mu_t}},\epsilon_{\mu_t}-1]}\cap K_s}
								\eta^{\mu_t-j-1}||\hat{y}_{j|\epsilon_{\mu_t}}^*-{y}_j||_R^2 \big)
							\end{aligned}	
							\label{eq:boundETcond}
						\end{align}
						where we used the fact that $\mu_t-M_{\mu_{t}}=\epsilon_{\mu_t}-M_{\epsilon_{\mu_t}}$.
						\par Now we want to upper bound $2\sum_{j\in\mathbb{I}_{[t-M_t,\mu_t-1]}\setminus K_s}\eta^{t-j-1} ||\hat{y}_{j|\mu_t}^*-\hat{y}_{j|t}^*||_R^2$, i.e., the second term on the right-hand side of (\ref{eq:boundDeltaY}). For $\mu_t \leq t-M_t$ the sum is empty  and thus $2\sum_{j\in\mathbb{I}_{[t-M_t,\mu_t-1]}\setminus K_s}\eta^{t-j-1} ||\hat{y}_{j|\mu_t}^*-\hat{y}_{j|t}^*||_R^2=0$. 
						\textcolor{rev}{
							For $\mu_t>t-M_t$
							notice that due to $\gamma_{\mu_t}=0$, $\mu_{t-1} \notin K_s$  and hence we can apply the same argument as in the proof of Proposition~\ref{prop:NLPOL} to conclude that 
							$\hat{y}^*_{j | \mu_t} = \hat{y}^*_{j | \mu_t-1}$ for $j \in \mathbb{I}_{[t-M_t,\mu_t-2]}$ and $\hat{w}^*_{\mu_t-1 | \mu_t} = 0$.} Thus, we obtain
						\begin{align*}
							\begin{aligned}
								&2\sum_{j\in\mathbb{I}_{[t-M_t,\mu_t-1]}\setminus K_s}\eta^{t-j-1} ||\hat{y}_{j|\mu_t}^*-\hat{y}_{j|t}^*||_R^2 \\&
								=2\eta^{t-\mu_t}
								\sum_{j\in\mathbb{I}_{[t-M_t,\mu_t-2]}\setminus K_s}
								\eta^{\mu_t-j-1} ||\hat{y}^*_{j |\mu_t-1}-\hat{y}_{j |t}^*||_R^2\\&+2\eta^{t-\mu_t} ||\hat{y}_{\mu_t-1 |t}^*-h(\hat{x}_{\mu_t-1},u_{\mu_t-1},0)||_R^2.
							\end{aligned}	
						\end{align*}
						Hence, by  constraint (\ref{eq:NLPC4}) we can write 
						\begin{align}
							\begin{aligned}
								&2\sum_{j\in\mathbb{I}_{[t-M_t,\mu_t-1]}\setminus K_s}\eta^{t-j-1} ||\hat{y}_{j|\mu_t}^*-\hat{y}_{j|t}^*||_R^2 \\&
								\leq
								2\eta^{t-\mu_t}\big(\alpha \sum_{j={\mu_t}-M_{\mu_t}}^{\epsilon_{\mu_t}-1}\eta^{\mu_t-j-1} 2||\hat{w}_{j|\epsilon_{\mu_t}}^*||_Q^2\\&+ \alpha \sum_{j\in \mathbb{I}_{[{\mu_t}-M_{\mu_t},\epsilon_{\mu_t}-1]}\cap K_s}\eta^{\mu_t-j-1}||\hat{y}_{j|\epsilon_{\mu_t}}^*-y_j||_R^2 \big).
							\end{aligned}	
							\label{eq:boundp2}
						\end{align}
						Substituting (\ref{eq:boundETcond}) and (\ref{eq:boundp2}) in (\ref{eq:boundDeltaY}) and exploiting Proposition~\ref{prop:NLPOL}, we obtain
						\begin{align}
							\begin{aligned}
								&\sum_{j\in\mathbb{I}_{[t-M_t,\mu_t-1]}\setminus K_s}\eta^{t-j-1}||y_j-\hat{y}_{j|t}^*||_R^2\\&\leq 4\eta^{t-\mu_t}\alpha\Big(\sum_{j=\mu_t-M_{\mu_t}}^{\mu_t-1} \eta^{\mu_t-j-1}2 ||\hat{w}^*_{j|\mu_t}||_Q^2\\& +\sum_{j\in \mathbb{I}_{[\mu_t-M_{\mu_t},\mu_t -1]}\cap K_s} \eta^{\mu_t-j-1} ||\hat{y}^*_{j|\mu_t}-y_j||_R^2 \Big)\\&
								\leq 4\eta^{t-\mu_t} J(\hat{x}_{\mu_t-M_{\mu_t}|\mu_t}^*, \hat{w}_{\cdot|\mu_t}^*, \hat{y}_{\cdot|\mu_t}^*,\mu_t)\\&\leq 4\eta^{t-\mu_t} J(x_{\mu_t-M_{\mu_t}}, w_{\cdot|\mu_t}, y_{\cdot|\mu_t},\mu_t)
							\end{aligned}
							\label{eq:Jepsilonbar}
						\end{align}
						where   $w_{\cdot|\tau}$ and  $y_{\cdot|\tau}$ refer to the true disturbance and output trajectories on the interval $[\tau-M_\tau, \tau-1]$ \color{new} and by optimality and Lemma~\ref{lem:addC} it holds that\footnote{By Lemma~\ref{lem:addC}, the true system trajectories satisfy all the constraints of the optimization problem, and hence are a valid candidate solution.} $J(\hat{x}_{\mu_t-M_{\mu_t}|\mu_t}^*, \hat{w}_{\cdot|\mu_t}^*, \hat{y}_{\cdot|\mu_t}^*,\mu_t)\leq  J(x_{\mu_t-M_{\mu_t}}, w_{\cdot|\mu_t}, y_{\cdot|\mu_t},\mu_t)$. \color{black} 
						Moreover, we can upper bound the following terms of (\ref{eq:ET2eiossEstsplit2}) as follows
							\begin{align*}
								\begin{aligned}
									&2\eta^{M_t}||\hat{x}_{t-M_t|t}^*-\hat{x}_{t-M_t}||_{P_2}^2+	\sum_{j=t-M_t}^{t-1}\eta^{t-j-1}2||\hat{w}_{j|t}^*||_Q^2\\&	 +\sum_{j\in \mathbb{I}_{[t-M_t,t-1]}\cap K_s}\eta^{t-j-1}||\hat{y}_{j|t}^*-y_j||_R^2\\\leq &J(\hat{x}_{t-M_t|t}^*, \hat{w}_{\cdot|t}^*, \hat{y}_{\cdot|t}^*,t).
								\end{aligned}
							\end{align*} 
							\color{new}
							Since $J(\hat{x}_{t-M_t|t}^*, \hat{w}_{\cdot|t}^*, \hat{y}_{\cdot|t}^*,t) \leq J(x_{t-M_t}, w_{\cdot|t}, y_{\cdot|t},t)$ again by optimality and Lemma~\ref{lem:addC}, we obtain  
								\begin{align}
								\begin{aligned}
									&2\eta^{M_t}||\hat{x}_{t-M_t|t}^*-\hat{x}_{t-M_t}||_{P_2}^2\\&+	\sum_{j=t-M_t}^{t-1}\eta^{t-j-1}2||\hat{w}_{j|t}^*||_Q^2\\&	 +\sum_{j\in \mathbb{I}_{[t-M_t,t-1]}\cap K_s}\eta^{t-j-1}||\hat{y}_{j|t}^*-y_j||_R^2\\\leq& J(x_{t-M_t}, w_{\cdot|t}, y_{\cdot|t},t).
								\end{aligned}
								\label{eq:Jalpha}
							\end{align}
							\color{black}
						Substituting (\ref{eq:Jepsilonbar}) and (\ref{eq:Jalpha})	in	(\ref{eq:ET2eiossEstsplit2}) results in
						\begin{align}
							\begin{aligned}
								||\hat{e}_t||_{P_1}^2&=||\hat{x}_t-x_t||_{P_1}^2 \\&\leq 4\eta^{M_t} ||\hat{x}_{t-M_t}-x_{t-M_t}||_{P_2}^2\\&+8\eta^{t-\mu_t+M_{\mu_t}} ||\hat{x}_{\mu_t-M_{\mu_t}}-x_{\mu_t-M_{\mu_t}}||_{P_2}^2\\&+\max\{2\alpha+2,4\}\sum_{j=t-M_t}^{t-1}\eta^{t-j-1} ||w_j||_Q^2\\&+\max\{8\alpha,8\}\sum_{j=\mu_t-M_{\mu_t}}^{\mu_t-1}\eta^{t-j-1} ||w_j||_Q^2\\
								&\leq 4\eta^{M_t} ||\hat{x}_{t-M_t}-x_{t-M_t}||_{P_2}^2\\&+8\eta^{t-\mu_t+M_{\mu_t}} ||\hat{x}_{\mu_t-M_{\mu_t}}-x_{\mu_t-M_{\mu_t}}||_{P_2}^2\\&+\max\{10\alpha+2,12\}\sum_{j=\mu_t-M_{\mu_t}}^{t-1}\eta^{t-j-1} ||w_j||_Q^2
								\\&\leq \max\{ 12 \eta^{M_t} ||\hat{x}_{t-M_t}-x_{t-M_t}||_{P_2}^2,\\&24\eta^{t-\mu_t+M_{\mu_t}} ||\hat{x}_{\mu_t-M_{\mu_t}}-x_{\mu_t-M_{\mu_t}}||_{P_2}^2,\\&3\max\{10\alpha+2,12\}\sum_{j=\mu_t-M_{\mu_t}}^{t-1}\eta^{t-j-1} ||w_j||_Q^2\}.
							\end{aligned}
							\label{eq:sameNvar}
						\end{align}
					As discussed below  (\ref{eq:ET2eiossEstsplit2}), we derived  (\ref{eq:sameNvar})  for $t\geq \tilde{\tau}$. For the case of $t<\tilde\tau$, note that the following provides a valid upper bound
							\begin{align}
									\label{eq:sameNvar2}
								\begin{aligned}
									||\hat{e}_t||_{P_1}^2&\leq \max\{ 12 \eta^{M_t} ||\hat{x}_{t-M_t}-x_{t-M_t}||_{P_2}^2,\\&3\max\{10\alpha+2,12\}\sum_{j=t-M_t}^{t-1}\eta^{t-j-1} ||w_j||_Q^2\}.
								\end{aligned}
							\end{align}	
						 	Due to $||\cdot||_{P_2}^2\leq\lambda_{\text{max}}(P_2,P_1)||\cdot||_{P_1}^2$, we obtain the following upper bound for (\ref{eq:sameNvar}) 
						\begin{align*}
							\begin{aligned}
								||&\hat{e}_t||_{P_1}^2\\
								\leq &\max\{ 12 \eta^{M_t} \lambda_{\text{max}}(P_2,P_1)||\hat{x}_{t-M_t}-x_{t-M_t}||_{P_1}^2,\\&24\eta^{t-\mu_t+M_{\mu_t}}\lambda_{\text{max}}(P_2,P_1) ||\hat{x}_{\mu_t-M_{\mu_t}}-x_{\mu_t-M_{\mu_t}}||_{P_1}^2,\\&3\max\{10\alpha+2,12\}\sum_{j=\mu_t-M_{\mu_t}}^{t-1}\eta^{t-j-1} ||w_j||_Q^2\}.
							\end{aligned}
						\end{align*}	
						Selecting the horizon length  $M$ large enough such that 
						\begin{align}
							\rho^M\coloneqq24\lambda_{\text{max}}(P_2,P_1)\eta^{M}<1
							\label{eq:minhor}
						\end{align}
						with $\rho \in [0,1)$, noting that $\rho\geq\eta$, and recalling that \mbox{$t-\mu_t+M_{\mu_t}\geq M_t \geq M$} for all $t\geq \max\{M,\tilde\tau\}$, we obtain for all $t\geq \max\{M,\tilde\tau\}$
						\begin{align}
							\begin{aligned}
								||\hat{e}_t||_{P_1}^2&\leq \max\{\rho^{M_t} ||\hat{x}_{t-M_t}-x_{t-M_t}||_{P_1}^2,\\&\rho^{t-\mu_t+M_{\mu_t}} ||\hat{x}_{\mu_t-M_{\mu_t}}-x_{\mu_t-M_{\mu_t}}||_{P_1}^2,\\ &3 \max\{10\alpha+2,12\}\sum_{j=\mu_t-M_{\mu_t}}^{t-1}\eta^{t-j-1} ||w_j||_Q^2 \}. 
							\end{aligned}
							\label{eq:ET2boundMstepET}
						\end{align}
						Notice that in case of $\gamma_t=0$, $\mu_t=t$ and $M_t=t-\mu_t+M_{\mu_t}$, thus the first two terms of the right-hand side of (\ref{eq:ET2boundMstepET}) are identical. 
						Consider some time $l\in \mathbb{I}_{[0,M-1]}$. %
						Using (\ref{eq:sameNvar}) if $l\geq\tilde\tau$ or (\ref{eq:sameNvar2}) if $l<\tilde\tau$, we can write
						\begin{align}
							\begin{aligned}
								||\hat{e}_l||_{P_1}^2&\leq \max\{12\eta^{l} ||\hat{e}_0||_{P_2}^2,24\eta^{l} ||\hat{e}_0||_{P_2}^2,\\ &3 \max\{10\alpha+2,12\}\sum_{j=0}^{l-1}\eta^{l-j-1} ||w_j||_Q^2 \}. 
							\end{aligned}
							\label{eq:ET2boundlessM}
						\end{align}
						Next we recursively apply the bound 	(\ref{eq:ET2boundMstepET}) to the first two terms on the right-hand side of 	(\ref{eq:ET2boundMstepET})\footnote{\textcolor{rev}{Similar steps using a sum-based formulation are followed in \cite[Corollary 1]{Sch23} and \cite[Theorem 1]{Kra24}.}}. 
						Applying (\ref{eq:ET2boundMstepET})  recursively, using (\ref{eq:ET2boundlessM}) when we reach a time instant less than $M$, and using that $\eta\leq \rho$, we obtain the following bound 
						\begin{align*}
							\begin{aligned}
								||\hat{e}_t||_{P_1}^2&\leq \max\{12\rho^{t} ||\hat{e}_0||_{P_2}^2,24\rho^{t} ||\hat{e}_0||_{P_2}^2\}\\&+3 \max\{10\alpha+2,12\}\sum_{j=0}^{t-1}\rho^{t-j-1} ||w_j||_Q^2. 
							\end{aligned}
						\end{align*}
						Notice that we upper bounded all disturbance terms that result from applying (\ref{eq:ET2boundMstepET}) recursively and (\ref{eq:ET2boundlessM}) by the sum of all disturbances over the whole time interval $[0,t-1]$.
						Since $\lambda_{\text{min}}(P_1)||\hat{e}_t||^2\leq ||\hat{e}_t||_{P_1}^2$, $||\hat{e}_0||_{P_2}^2\leq \lambda_{\text{max}}(P_2)||\hat{e}_0||^2$, and   $||w_j||_{Q}^2\leq \lambda_{\text{max}}(Q)||w_j||^2$ it holds that
						\begin{align*}
							\begin{aligned}
								||\hat{e}_t||^2&\leq  24 \frac{\lambda_{\text{max}}(P_2)}{\lambda_{\text{min}}(P_1)} \rho^{t}||\hat{e}_{0}||^2\\&+\frac{3\max\{10\alpha+2,12\}\lambda_{\text{max}}(Q)}{\lambda_{\text{min}}(P_1)}\sum_{j=0}^{t-1} \rho^{t-j-1}  ||w_j||^2.
							\end{aligned}
						\end{align*}	
						Using $\sqrt{a+b}\leq \sqrt{a}+\sqrt{b}$ for all $a,b\geq0$ results in 
						\begin{align*}
							\begin{aligned}
								||\hat{e}_t||&\leq	 \sqrt{\frac{24\lambda_{\text{max}}(P_2)}{\lambda_{\text{min}}(P_1)}}\sqrt{\rho}^{t}||\hat{e}_{0}|| \\&+\sqrt{\frac{3\max\{10\alpha+2,12\}\lambda_{\text{max}}(Q)}{\lambda_{\text{min}}(P_1)}}\sum_{j=0}^{t-1} \sqrt{\rho}^{t-j-1}  ||w_j||
							\end{aligned}
						\end{align*}
						and concludes the proof.	
					\end{proof}
					\par	Theorem~\ref{thm:stab2} shows that the proposed ET-MHE scheme is an RGES estimator. We note that the disturbance gain increases with increasing $\alpha$. This is to be expected since for larger $\alpha$, events are triggered less often. On the other hand, the condition on the required horizon length is independent of $\alpha$. This is crucial since this means that the MHE scheme does not need to be re-designed if the triggering parameter $\alpha$ is changed (see also Remark~\ref{rem:alphat} below). We note that $\alpha$ could also be omitted in the cost function (8). However, then the RGES proof needs to be suitably adapted, which would result in an increase of the minimum horizon length for larger values of $\alpha$. By including $\alpha$ in the cost function, we obtain a minimum horizon length independent of $\alpha$ as shown in the proof of Theorem~\ref{thm:stab2}.		\color{resl} In particular, note that due to the choice of $J$, the factor in front of $J$ in (\ref{eq:Jepsilonbar}) is independent of $\alpha$, and thus the resulting condition on the minimum horizon length is likewise independent of $\alpha$. \color{black}
					\begin{rem}
						\label{rem:alphat}
						The choice of $\alpha$ is an important consideration for the performance of the proposed estimator. Although we do not explore this further in this paper, it is possible to use a time-varying $\alpha_t$ in the event triggering condition. This could be advantageous because this  enables adjusting the sensitivity of the event-triggering condition. For example, it can help reducing excessive triggering at the start due to a poor prior. When using a time-varying $\alpha_t$, the  cost function must be adjusted accordingly. Either a time varying $\alpha_t$ must also be used in the cost function, or  alternatively, the cost function can be formulated using the largest value of $\alpha_t$ that will be used at any point in time. This largest value of $\alpha_t$  then determines the bound on the estimation error.
					\end{rem}
					\section{ET-MHE with varying horizon length}
					\label{sec:varN}
					\par In the following, we present a modified version of the ET-MHE scheme proposed in Section~\ref{sec:MHEScheme}. While the previous NLP (\ref{eq:ET2ETNLP2})-(\ref{eq:objfunc2}) was always explicitly solved with a fixed horizon of length $M$ for $t \geq M$ (compare Remark~\ref{rem:M}),  the following scheme allows to vary the horizon length every time the optimization problem is solved. We show that this allows to establish a tighter bound on the estimation error at the cost of higher computational complexity and a less intuitive set up. This modified ET-MHE scheme uses the same algorithm as Algorithm~\ref{alg2}. The only differences between the two methods lie in the definition of the horizon length and the formulation of the cost function, both of which we introduce in the following.			
					\par 	Before specifying the horizon length $M_t$, it is essential to define  $\sigma_t=\max\{\tau\leq t|\gamma_j=1, \forall j\in[\tau-\min\{\tau,2M-1\},\tau]\}$\footnote{\textcolor{rev}{Note that in Algorithm~\ref{alg2}, $\gamma_0=1$, and thus $\sigma_t$ is well defined for all $t\geq 0$.}}, i.e., the most recent time at which $\min\{t,2M\}$ consecutive events had occurred. 
						We define the horizon length as \textcolor{rev}{\footnote{\textcolor{rev}{Note that if $t-\delta_{t}-M<\tilde\tau$, then $\mu_{t-\delta_{t}-M}$ is not defined and the definition of $M_t$ reduces to $M_t\coloneqq \min\{t,t-\sigma_t+M\}$.}}}
						\begin{align}
							M_t\coloneqq \min\{t,t-\mu_{t-\delta_t-M},t-\sigma_t+M\}.
							\label{eq:Mtnew22}
						\end{align}
						The main idea here is as follows. If there is an event at time $t$, we do not solve the NLP with a fixed horizon $M$. Instead, the horizon depends on whether there was an event at time $t-M$. If this is the case, we extend the time interval considered in the cost function from $[t-M,t-1]$ to $[\mu_{t-M},t-1]$. Thereby, if multiple consecutive measurements had been transmitted prior to $t-M$, they are additionally taken into account in the optimization problem. 
						The inclusion of  $\sigma_t$ in the definition of the horizon length serves to ensure an upper bound on the maximum potential horizon length even in scenarios with an arbitrarily large number of consecutive events. By  (\ref{eq:Mtnew22}), less than $2M$ consecutive measurements are considered  in the cost function, thus constraining the horizon length to $M_t<3M$.
						Although such a scenario is not typically expected, $\sigma_t$ is incorporated in the horizon length definition to guarantee that it will not become unbounded and ever-increasing.
						This choice of horizon length allows for the establishment of a tighter bound on the estimation error and a smaller minimum horizon length as we show below.

						Furthermore, the cost function (\ref{eq:objfunc2}) is modified by replacing $\max\{\alpha,1\}$ with $(\alpha+1)$, i.e., the cost function is now given by
						\begin{align}
							\begin{aligned}
								&J(\hat{x}_{t-M_t|t}, \hat{w}_{\cdot|t},  \hat{y}_{\cdot|t},t)\\=&2\eta^{M_t}||\hat{x}_{t-M_t|t}-\hat{x}_{t-M_t}||_{P_2}^2 \\+&(\alpha+1)\Big(\sum_{j=t-M_t}^{t-1} \eta^{t-j-1}2 ||\hat{w}_{j|t}||_Q^2\\+&\sum_{j\in \mathbb{I}_{[t-M_t,t-1]}\cap K_s} \eta^{t-j-1} ||\hat{y}_{j|t}-y_j||_R^2 \Big).
							\end{aligned}
							\label{eq:objfunc3}
						\end{align}
						This modification allows us  to exploit  the varying horizon length to obtain a tighter error bound. 
						Before stating the corresponding theorem, we introduce first the following lemma that will be used
						to prove RGES.
						\begin{lem}
							\label{lem:bound_mu}
							Let Assumption \ref{ass:eIOSSET} hold. Consider the ET-MHE
							scheme~(\ref{eq:ET2ETNLP2})-(\ref{eq:NLPC4}), and (\ref{eq:objfunc3}) with the event-triggering condition (\ref{eq:ET2trigruledAdap}) and with $M_t$ as defined in (\ref{eq:Mtnew22}). Then,  for all $t\geq 0$, if either  $\gamma_t=0$ or  $\gamma_j=1, \ \forall j\in[t-\min\{t,M-1\},t]$, 
							the following holds
							\begin{align}
								\begin{aligned}
									||\hat{x}_{t}-x_{t}||_{P_1}^2 &\leq 4\eta^{M_{t}} ||\hat{x}_{{t}-M_{t}}-x_{{t}-M_{t}}||_{P_2}^2\\&+(2\alpha+4)\sum_{j={t}-M_{t}}^{{t}-1}\eta^{{t}-j-1} ||w_j||_Q^2.
								\end{aligned}
								\label{eq:ET2boundMtstep1ET}
							\end{align} 
						\end{lem}
						\begin{proof}
							We can apply the same steps as in the proof  of Theorem~\ref{thm:stab2} to obtain (\ref{eq:ET2eiossEstsplit2}). 
							Now, first consider the case of  $\gamma_t=0$. 	If $\gamma_t=0$, i.e,   the bound in (\ref{eq:ET2trigruledAdap}) holds, then the bound in (\ref{eq:ET2trigruled2}) also holds. Note also, that, by (\ref{eq:Mtnew22}), if $\gamma_t=0$ then $t-M_t = \epsilon_t-M_{\epsilon_t}$. Hence, using Proposition~\ref{prop:NLPOL}, we can upper bound the last term in (\ref{eq:ET2eiossEstsplit2}) as follows
							\begin{align}
								\begin{aligned}
									&\sum_{j\in\mathbb{I}_{[t-M_{t},t-1]}\setminus K_s}\eta^{t-j-1} ||y_j-\hat{y}_{j|t}^*||_R^2 \\<& \alpha \Big( 2\sum_{j={t}-M_{t}}^{\epsilon_{t}-1} \eta^{{t}-1-j} ||\hat{w}_{j|\epsilon_{t}}^*||_Q^2\\&+\sum_{j\in\mathbb{I}_{[t-M_t,\epsilon_{t}-1]}\cap K_s} \eta^{{t}-1-j}||\hat{y}_{j|\epsilon_{t}}^*-y_j||_R^2\Big)
								\end{aligned}
								\label{boundbyETM}
							\end{align}
							resulting in 
							\begin{align*}
								\begin{aligned}
									||\hat{x}_{t}-x_{t}||_{P_1}^2 &\leq 2\eta^{M_{t}}||\hat{x}_{{t}-M_{t}}-x_{{t}-M_{t}}||_{P_2}^2\\&+2\eta^{M_{t}}||\hat{x}_{{t}-M_{t}|{t}}^*-\hat{x}_{{t}-M_{t}}||_{P_2}^2\\&+\sum_{j={t}-M_{t}}^{{t}-1}\eta^{{t}-j-1} 2||w_j||_Q^2\\
									&+(\alpha+1)\big(\sum_{j={t}-M_{t}}^{{t}-\delta_{t}-1}\eta^{{t}-j-1} 2 ||\hat{w}_{j|{t}-\delta_{t}}^*||_Q^2\\ 
									&+\sum_{j\in \mathbb{I}_{[t-M_t,t-\delta_t-1]}\cap K_s}\eta^{{t}-j-1}||\hat{y}_{j|{t}-\delta_{t}}^*-y_j||_R^2\big).
								\end{aligned}
							\end{align*}
							Recalling the choice of our cost function, we can equivalently write
							\begin{align*}
								\begin{aligned}
									||\hat{x}_{t}-x_{t}||_{P_1}^2 &\leq 2 \eta^{M_{t}} ||\hat{x}_{{t}-M_{t}}-x_{{t}-M_{t}}||_{P_2}^2\\&+\sum_{j={t}-M_t}^{{t}-1}2\eta^{{t}-j-1} ||w_j||_Q^2 \\&+J(\hat{x}_{{t}-M_{t}|{t}}^*, \hat{w}_{\cdot|{t}}^*,  \hat{y}_{\cdot|{t}}^*,{t}) .
								\end{aligned}
							\end{align*}
							Since $J(\hat{x}_{t-M_{t}|{t}}^*, \hat{w}_{\cdot|{t}}^*, \hat{y}_{\cdot|{t}}^*,{t}) \leq J(x_{{t}-M_{t}}, w_{\cdot|{t}}, y_{\cdot|{t}},{t})$	by optimality, the inequality in (\ref{eq:ET2boundMtstep1ET}) holds as desired.
							\par 	If $\gamma_j=1, \ \forall j\in[t-\min\{t,M-1\},t]$, then by (\ref{eq:Mtnew22})  
							the optimization problem is solved over a horizon 	$M_t<2M$ with measurements available at every time instant in the horizon, i.e., 
							$\tau \in K_s$ for all $\tau\in [t-M_t,t-1]$. (Recall that if $\gamma_\tau=1$ then $y_{\tau-1}$ is transmitted.) 
							Hence, the last term of (\ref{eq:ET2eiossEstsplit2}) is zero and we can directly obtain (\ref{eq:ET2boundMtstep1ET}) without the intermediate step (\ref{boundbyETM}).		
						\end{proof}
						\begin{thm}[RGES of ET-MHE  with varying $M_t$]
								\label{thm:stab3}
							Let Assumption~\ref{ass:eIOSSET} hold and $M\in \mathbb{I}_{\geq 0}$ be chosen such that $8\lambda_{\text{max}}(P_2,P_1)\eta^{M}<1$. Then there exists $\rho \in [0,1)$ such that the state estimation error of the ET-MHE
							scheme~(\ref{eq:ET2ETNLP2})-(\ref{eq:NLPC4}) and (\ref{eq:objfunc3}) with the even-triggering condition (\ref{eq:ET2trigruledAdap}) and with $M_t$ as defined in (\ref{eq:Mtnew22}) satisfies for all $t\geq 0$ 
							\begin{align}
								\begin{aligned}
									||\hat{e}_t||&\leq	 \sqrt{\frac{8\lambda_{\text{max}}(P_2)}{\lambda_{\text{min}}(P_1)}}\sqrt{\rho}^{t}||\hat{e}_{0}|| \\&+\sqrt{\frac{(10\alpha+12)\lambda_{\text{max}}(Q)}{\lambda_{\text{min}}(P_1)}}\sum_{j=0}^{t-1} \sqrt{\rho}^{t-j-1}  ||w_j||,
								\end{aligned}
								\label{eq:ET2errorboundnew}
							\end{align}
							i.e., the ET-MHE is an RGES estimator according to Definition~\ref{def:rges}.
						\end{thm}
						\begin{proof}
							We can apply the same arguments from the proof of Theorem~\ref{thm:stab2} up to (\ref{eq:sameNvar}), to obtain
							\begin{align*}
								\begin{aligned}
									||\hat{e}_t||_{P_1}^2&\leq 4\eta^{M_t} ||\hat{x}_{t-M_t}-x_{t-M_t}||_{P_2}^2\\&+8\eta^{t-\mu_t+M_{\mu_t}} ||\hat{x}_{\mu_t-M_{\mu_t}}-x_{\mu_t-M_{\mu_t}}||_{P_2}^2\\&+(2\alpha+4)\sum_{j=t-M_t}^{t-1}\eta^{t-j-1} ||w_j||_Q^2\\&+(8\alpha+8)\sum_{j=\mu_t-M_{\mu_t}}^{\mu_t-1}\eta^{t-j-1} ||w_j||_Q^2.
								\end{aligned}
							\end{align*}
							Consider some time $t>M$ for some  $M$ such that 
							\begin{align}
								\rho^M\coloneqq 8\lambda_{\text{max}}(P_2,P_1)\eta^{M}<1
								\label{eq:minhornew}
							\end{align}
							with $\rho \in [0,1)$.	
							Due to  $||\hat{e}_{\tau-M_{\tau}}||_{P_2}^2\leq \lambda_{\text{max}}(P_2,P_1)||\hat{e}_{\tau-M_{\tau}}||_{P_1}^2$,  (\ref{eq:minhornew}) and $\rho\geq\eta$, we can write
							\begin{align}
								\begin{aligned}
									||\hat{e}_t||_{P_1}^2 
									&\leq \rho^{M_t} ||\hat{e}_{t-M_t}||_{P_1}^2+\rho^{t-\mu_t+M_{\mu_t}} ||\hat{e}_{\mu_t-M_{\mu_t}}||_{P_1}^2\\&+(2\alpha+4)\sum_{j=t-M_t}^{t-1}\eta^{t-j-1} ||w_j||_Q^2\\&+(8\alpha+8)\sum_{j=\mu_t-M_{\mu_t}}^{\mu_t-1}\eta^{t-j-1} ||w_j||_Q^2.
								\end{aligned}						\label{eq:etbound2}										\end{align}
							By definition of $M_t$ in (\ref{eq:Mtnew22}) we know that either $\gamma_{t-M_t}=0$ or  $\gamma_j=1, \ \forall j\in[t-M_t-\min\{t-M_t,M-1\},t-M_t]$ and, moreover, that either  $\gamma_{\mu_t-M_{\mu_t}}=0$  or $\gamma_j=1, \ \forall j\in[\mu_t-M_{\mu_t}-\min\{\mu_t-M_{\mu_t},M-1\},\mu_t-M_{\mu_t}]$, and thus Lemma~\ref{lem:bound_mu} is applicable. Using again $||\hat{e}_{\tau-M_{\tau}}||_{P_2}^2\leq \lambda_{\text{max}}(P_2,P_1)||\hat{e}_{\tau-M_{\tau}}||_{P_1}^2$,  (\ref{eq:minhornew}) and $\rho\geq\eta$,
							(\ref{eq:ET2boundMtstep1ET}) can be upper bounded for  $\tau\geq M$ by
							\begin{align}
								\begin{aligned}
									||\hat{x}_{\tau}-x_{\tau}||_{P_1}^2 &\leq \rho^{M_{\tau}} ||\hat{x}_{{\tau}-M_{\tau}}-x_{\tau-M_\tau}||_{P_1}^2\\&+(2\alpha+4)\sum_{j=\tau-M_\tau}^{\tau-1}\rho^{{\tau}-j-1} ||w_j||_Q^2.
								\end{aligned}
								\label{eq:ET2boundMtstep1ET2}
							\end{align}		
							By recursively applying  (\ref{eq:ET2boundMtstep1ET2}) to the first two terms on the right-hand side of (\ref{eq:etbound2}), using (\ref{eq:ET2boundMtstep1ET})  upon reaching a time instance less than $M$,   and considering that $\eta\leq \rho$, 
							we derive the following bound
							\begin{align*}
								\begin{aligned}
									||\hat{e}_t||_{P_1}^2&\leq 8\rho^{t} ||\hat{e}_0||_{P_2}^2+ (10\alpha+12)\sum_{j=0}^{t-1}\rho^{t-j-1} ||w_j||_Q^2. 
								\end{aligned}
							\end{align*}
							Performing reformulations analogously to the steps in the proof of Theorem \ref{thm:stab2}, we finally obtain (\ref{eq:ET2errorboundnew}).
						\end{proof}
						\textcolor{rev}{Due to the new definition of the horizon length in (\ref{eq:Mtnew22})  the bound derived  in Lemma~\ref{lem:bound_mu} can be recursively applied in the proof above. 
							This allows to transition from (\ref{eq:etbound2}) to (\ref{eq:ET2boundMtstep1ET2}) without the need to use a max-based bound as in the proof of Theorem~\ref{thm:stab2} (compare the last inequality in (\ref{eq:sameNvar})). This results in a smaller error bound and minimum horizon length.}
					\par The discussion following the proof of Theorem~\ref{thm:stab2} 
					remains valid for this modified  ET-MHE scheme with varying horizon length.
					\par While extending the ET-MHE scheme from Section~\ref{sec:MHEScheme} to allow a varying horizon length enables the derivation of a smaller error bound and a smaller minimal required horizon length (compare Theorem~\ref{thm:stab3} with  Theorem~\ref{thm:stab2}), the original scheme in Section~\ref{sec:MHEScheme} remains simpler to implement and computationally less demanding. This is due to the use of the fixed horizon length, which avoids the need to determine the horizon length at each event and allows the optimization problem to retain a constant structure throughout.
					\section{Numerical Examples}
					\label{sec:sim}
					In this section we present simulation examples for the ET-MHE scheme from Section~\ref{sec:MHEScheme}. \color{resl} Additionally, we illustrate some differences between the proposed method and the one form our previous work \cite{Kra24}. \color{black} Furthermore, we also discuss the influence of the additional constraint (\ref{eq:NLPC4})  on the estimation results and  the modified  scheme with varying horizon length. 
					\subsection{Example 1: Batch reactor}
					In the first example consider the following system
					\begin{align*}
						\begin{aligned}
							x_{1,t+1}&=x_{1,t}+\tau(-2k_1x_{1,t}^2+2k_2x_{2,t})+w_{1,t},\\
							x_{2,t+1}&=x_{2,t}+\tau(k_1x_{1,t}^2-k_2x_{2,t})+w_{2,t},\\
							y&=x_{1,t}+x_{2,t}+w_{3,t}
						\end{aligned}
					\end{align*}
					with $k_1=0.16,\ k_2=0.0064$, and sampling time $\tau=0.1$. This corresponds to a  batch reactor system from \cite{Ten02}, where the state $x$ represents the partial pressures of the individual components in the batch reactor, discretized using the Euler method.
					This example has become a standard benchmark in the MHE literature (e.g. in \cite{Raw22,Sch23}), since other nonlinear state estimators such as the extended Kalman filter (EKF) often fail to yield meaningful results.
					For example, \cite[Example 4.38]{Raw22},  shows that with the same initial conditions as used here, the EKF, even when using measurements at every time step (i.e., no event-triggering), produces negative pressure estimates which leads to large estimation errors throughout the simulation. \color{black}
					We consider \mbox{$x_0=[3,1]^\top$} and the poor initial estimate $\hat{x}_0=[0.1,4.5]^\top$. In the simulation, the  additional disturbance $w\in \mathbb{R}^3$ is treated as  a uniformly distributed random variable that satisfies $|w_i|\leq 10^{-3}, \ i=1,2$, and $|w_3| \leq 0.1$. 
					\par For parameterizing the cost function, we use the parameters of the quadratic  i-IOSS Lyapunov function $W_{\delta}=||x-\tilde{x}||_P^2$ computed in \cite{Sch23}  where 
					\begin{align*}
						P=\begin{bmatrix}4.539&4.171\\4.171&3.834
						\end{bmatrix},
						\ Q=\begin{bmatrix}10^3 & 0&0\\0& 10^4 &0\\0 & 0& 10^3 \end{bmatrix}, \ R=10^3,	
					\end{align*}
					$P_1=P_2=P$ and the decay rate $\eta=0.91$. 
					These parameters satisfy Assumption \ref{ass:eIOSSET}.
					Using condition (\ref{eq:minhor}) we obtain a minimal horizon length of  $M_{\text{min}}=34$ guaranteeing robust stability of the ET-MHE. 
					\begin{figure}[!t]
						\centering
%
%
\definecolor{mycolor1}{rgb}{0.46600,0.67400,0.18800}%
\definecolor{mycolor2}{rgb}{0.30100,0.74500,0.93300}%
\definecolor{dgreen}{rgb}{0,0.5,0.00000}%
\begin{tikzpicture}

\begin{axis}[%
	width=0.99\figurewidth,
	height=0.8\figurewidth,
	scale only axis,
	xmin=0,
	xmax=60,
	xlabel style={font=\color{white!15!black}},
	xlabel={$t$},
	ymin=0,
	ymax=4.5,
	ylabel style={font=\color{white!15!black},at={(axis description cs:0.05,0.5)}},
	ylabel={$x,\hat{x}$},
axis background/.style={fill=white},
legend columns=2,
legend style={legend cell align=left, align=left, draw=white!15!black, font=\footnotesize}
]
\addplot [color=red, line width=1pt]
  table[row sep=crcr]{%
0	3\\
1	2.71420258703807\\
2	2.48056274849646\\
3	2.2843258429752\\
4	2.11811349306493\\
5	1.97639591080306\\
6	1.85313957195956\\
7	1.74494557890341\\
8	1.65046221491354\\
9	1.56470512372151\\
10	1.48826376145342\\
11	1.42025362673261\\
12	1.35836295240433\\
13	1.30242736241747\\
14	1.25067537707445\\
15	1.20354348748631\\
16	1.1605075057985\\
17	1.12001287226652\\
18	1.08259679478918\\
19	1.04683117125432\\
20	1.01375651385539\\
21	0.984097063803257\\
22	0.955645225514939\\
23	0.929126991994713\\
24	0.903360305681156\\
25	0.88077279563705\\
26	0.858886990010879\\
27	0.837898923022011\\
28	0.818293182679988\\
29	0.799070638819839\\
30	0.781894936690015\\
31	0.765128470874211\\
32	0.749641685258585\\
33	0.734127255307788\\
34	0.719294206556345\\
35	0.705384996426983\\
36	0.691574094442934\\
37	0.678750602714356\\
38	0.666597516523525\\
39	0.65607525226536\\
40	0.64495810314747\\
41	0.633780002012912\\
42	0.623873487305456\\
43	0.614457194810525\\
44	0.605997367623298\\
45	0.596830876607599\\
46	0.58819921536185\\
47	0.58033910990158\\
48	0.572266158441347\\
49	0.564129432079539\\
50	0.557465696833369\\
51	0.550190075056943\\
52	0.543775735945107\\
53	0.536713090718877\\
54	0.530741787516391\\
55	0.524442670879186\\
56	0.519147099756256\\
57	0.513602869263026\\
58	0.508325971250324\\
59	0.50220228564259\\
60	0.496221767754984\\
};
\addlegendentry{$x_1$}

\addplot [color=blue, line width=1pt]
  table[row sep=crcr]{%
0	0.1\\
1	3.44194350526519\\
2	2.59307576806921\\
3	2.35634639630895\\
4	2.26663651699362\\
5	2.02079566685958\\
6	1.89199659246932\\
7	1.70804697477764\\
8	1.6168839579901\\
9	1.53547882903543\\
10	1.46233759133736\\
11	1.3962596121395\\
12	1.3815054210645\\
13	1.31094690488029\\
14	1.25830818039788\\
15	1.21003089863702\\
16	1.15755002527163\\
17	1.11716659763218\\
18	1.07296089705341\\
19	1.04476719893499\\
20	1.01607946330645\\
21	0.98559354542309\\
22	0.957079869374323\\
23	0.930357010411606\\
24	0.90526526078389\\
25	0.881663457058013\\
26	0.859426350194381\\
27	0.838442412682439\\
28	0.818611999995263\\
29	0.799845801103359\\
30	0.781152344993577\\
31	0.769584121891104\\
32	0.753365871254188\\
33	0.736089017530011\\
34	0.723862790918155\\
35	0.712155178032235\\
36	0.698659234022975\\
37	0.685781216131883\\
38	0.673481963336601\\
39	0.661725555549816\\
40	0.650478983674318\\
41	0.639711859290125\\
42	0.629396158514674\\
43	0.619505995421749\\
44	0.610017421104869\\
45	0.600908245053537\\
46	0.592157875997377\\
47	0.583747179781177\\
48	0.575658352177059\\
49	0.567874804829688\\
50	0.560381062775686\\
51	0.553162672186768\\
52	0.548294612127791\\
53	0.519407928995875\\
54	0.513671032092157\\
55	0.508127460162239\\
56	0.502768697348425\\
57	0.497586713030413\\
58	0.492573927615933\\
59	0.487723181194275\\
60	0.483027704776495\\
};
\addlegendentry{$\hat{x}_{1,\alpha=5}$}

\addplot [color=red, dashdotted, line width=1pt]
  table[row sep=crcr]{%
0	1\\
1	1.14237338844413\\
2	1.25895547467407\\
3	1.35601883226001\\
4	1.43940256237984\\
5	1.51019989800811\\
6	1.5722763629508\\
7	1.62686311129136\\
8	1.67406187265392\\
9	1.71598375760948\\
10	1.75337655629286\\
11	1.78849914580933\\
12	1.81966307830149\\
13	1.84709453825341\\
14	1.8738406645462\\
15	1.89772237940677\\
16	1.91991758060288\\
17	1.94057461326621\\
18	1.96026046135205\\
19	1.97705943757123\\
20	1.99412198378726\\
21	2.01022656313591\\
22	2.02491298568684\\
23	2.03868059661974\\
24	2.05084084200242\\
25	2.06207101701469\\
26	2.07402289987279\\
27	2.08368647240332\\
28	2.09311386927238\\
29	2.10273617726919\\
30	2.1109173409985\\
31	2.11918322796974\\
32	2.12687609004076\\
33	2.13489049369761\\
34	2.1422348077122\\
35	2.14959425313987\\
36	2.15520370369857\\
37	2.16235874494061\\
38	2.16769198452633\\
39	2.17295350448944\\
40	2.17866188839261\\
41	2.18349448100347\\
42	2.18781178713426\\
43	2.19249062313476\\
44	2.19783629138657\\
45	2.20136885228893\\
46	2.20536563111121\\
47	2.20940789649296\\
48	2.21399230707558\\
49	2.21747246585737\\
50	2.22191974297289\\
51	2.22590632050309\\
52	2.23001677918525\\
53	2.23287705297731\\
54	2.23530292884531\\
55	2.23856714734476\\
56	2.24163490802197\\
57	2.24527071762472\\
58	2.24842757768932\\
59	2.25131774423731\\
60	2.25371823903535\\
};
\addlegendentry{$x_2$}

\addplot [color=blue, dashdotted, line width=1pt]
  table[row sep=crcr]{%
0	4.5\\
1	0.348072696215189\\
2	1.155766914834\\
3	1.29431271045372\\
4	1.25457254931668\\
5	1.46610131256902\\
6	1.53050084976415\\
7	1.71450483591179\\
8	1.76008634430556\\
9	1.8007889087829\\
10	1.83735952763193\\
11	1.87039851723086\\
12	1.77935199452044\\
13	1.84054118349516\\
14	1.86686054573637\\
15	1.8909991866168\\
16	1.94849868184088\\
17	1.9686903956606\\
18	2.01736661996654\\
19	2.00848445722349\\
20	1.99331354745033\\
21	2.00855650639201\\
22	2.0228133444164\\
23	2.03617477389776\\
24	2.04872064871161\\
25	2.06052155057455\\
26	2.07164010400637\\
27	2.08213207276234\\
28	2.09204727910593\\
29	2.10143037855188\\
30	2.16389001720314\\
31	2.13598470658175\\
32	2.14409383190021\\
33	2.1776180743641\\
34	2.16232122506281\\
35	2.1354186827191\\
36	2.14216665472373\\
37	2.14860566366928\\
38	2.15475529006692\\
39	2.16063349396031\\
40	2.16625677989806\\
41	2.17164034209016\\
42	2.17679819247788\\
43	2.18174327402434\\
44	2.18648756118278\\
45	2.19104214920845\\
46	2.19541733373653\\
47	2.19962268184463\\
48	2.20366709564669\\
49	2.20755886932038\\
50	2.21130574034738\\
51	2.21491493564184\\
52	2.1748433414127\\
53	2.26266421156483\\
54	2.26553266001669\\
55	2.26830444598165\\
56	2.27098382738855\\
57	2.27357481954756\\
58	2.2760812122548\\
59	2.27850658546563\\
60	2.28085432367452\\
};
\addlegendentry{$\hat{x}_{2,\alpha=5}$}

\addplot [color=dgreen, line width=1pt]
  table[row sep=crcr]{%
0	0.1\\
1	3.44344086305724\\
2	2.63423608954044\\
3	2.37339105484219\\
4	2.27283313760862\\
5	2.02568924219391\\
6	1.87752119979759\\
7	1.71959620948305\\
8	1.60511715424277\\
9	1.56752261427134\\
10	1.47740077786063\\
11	1.40850155508643\\
12	1.36426136200338\\
13	1.30591501667398\\
14	1.25420308041112\\
15	1.20545575308037\\
16	1.15885546968838\\
17	1.12344283322251\\
18	1.08308002036423\\
19	1.04818598130718\\
20	1.018324139747\\
21	0.989300514426997\\
22	0.955088507990684\\
23	0.928238538377136\\
24	0.903687872796883\\
25	0.88107229401101\\
26	0.858478025258885\\
27	0.837906231262037\\
28	0.816547827550239\\
29	0.797371362782897\\
30	0.781434862671077\\
31	0.765352286138569\\
32	0.745943419868796\\
33	0.732690020735196\\
34	0.71814836765979\\
35	0.7032871885985\\
36	0.692596256846986\\
37	0.680253154942028\\
38	0.676180998765722\\
39	0.670220040960696\\
40	0.661409776657173\\
41	0.64341610923086\\
42	0.62811057618132\\
43	0.630695930678588\\
44	0.611552316890735\\
45	0.602701703238255\\
46	0.608309626761098\\
47	0.601228572500069\\
48	0.595056338470557\\
49	0.578577090443772\\
50	0.555748026757284\\
51	0.560245611964858\\
52	0.561354061748267\\
53	0.529917456872952\\
54	0.525552053240831\\
55	0.509194504424547\\
56	0.497601229854908\\
57	0.489117989311634\\
58	0.48315487593318\\
59	0.479167183471241\\
60	0.471935444039023\\
};
\addlegendentry{$\hat{x}_{1, \alpha=0}$}

\addplot [color=dgreen, dashdotted, line width=1pt]
  table[row sep=crcr]{%
0	4.5\\
1	0.348325181075916\\
2	1.10826165666664\\
3	1.2743527433463\\
4	1.24695255852809\\
5	1.46000129961019\\
6	1.55729643854567\\
7	1.6955160649484\\
8	1.78299891206261\\
9	1.73638360274299\\
10	1.80496633913657\\
11	1.8437190920389\\
12	1.82944393717708\\
13	1.86268819961497\\
14	1.88509078961438\\
15	1.91893476329766\\
16	1.95237236062359\\
17	1.94234236577044\\
18	1.98278501964002\\
19	1.99094284128961\\
20	1.98635053111033\\
21	2.00639488609075\\
22	2.04317370126894\\
23	2.05321082831569\\
24	2.07834223316622\\
25	2.06789303687746\\
26	2.07709500443466\\
27	2.09623854457321\\
28	2.09379094152706\\
29	2.10022738274826\\
30	2.12881439809593\\
31	2.12629163248258\\
32	2.11892103867202\\
33	2.14454997640694\\
34	2.14458753629406\\
35	2.13470493973833\\
36	2.13963380686701\\
37	2.14325468816862\\
38	2.13366471344045\\
39	2.12077053742549\\
40	2.12985996488493\\
41	2.15592313186154\\
42	2.16857988875679\\
43	2.15307379931312\\
44	2.1815221064992\\
45	2.17989249001291\\
46	2.16180245661722\\
47	2.15540719332716\\
48	2.15320170009875\\
49	2.18278393509501\\
50	2.21307089009424\\
51	2.2005475964157\\
52	2.18647845118834\\
53	2.23820348948241\\
54	2.24513410202789\\
55	2.26673879693797\\
56	2.27209402026074\\
57	2.28090257794957\\
58	2.27246719841837\\
59	2.28053043510293\\
60	2.28648950029486\\
};
\addlegendentry{$\hat{x}_{2, \alpha=0}$}

\end{axis}

\end{tikzpicture}%
							\caption{Comparison of ET-MHE results for $\alpha=5$, MHE estimates (without event-triggering, i.e., $\alpha=0$) and real system states.} 
							\label{fig:etmhe2}
						\end{figure}
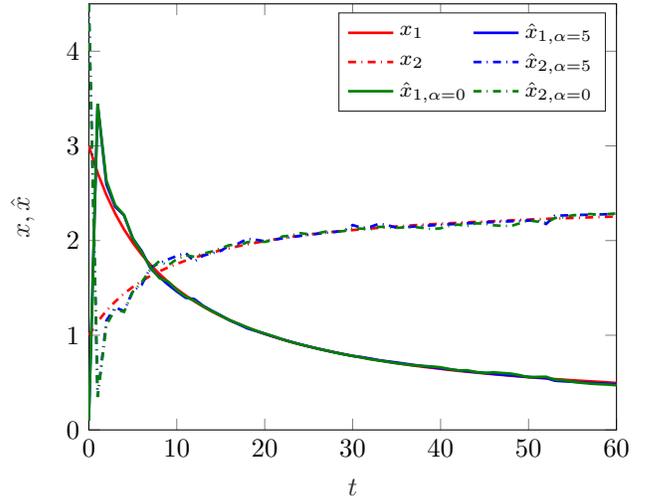
						
\begin{figure}[!t]
							\centering
%
%
\definecolor{mycolor1}{rgb}{0.00000,0.44700,0.74100}%
\begin{tikzpicture}

\begin{axis}[%
width=0.99\figurewidth,
height=0.3\figurewidth,
scale only axis,
xmin=0,
xmax=60,
xlabel style={font=\color{white!15!black}},
ytick={0,1},
xlabel={t},
ymin=0,
ymax=1,
ylabel style={font=\color{white!15!black},at={(axis description cs:0.03,0.5)}},
ylabel={$\gamma$},
axis background/.style={fill=white}
]
\addplot [color=mycolor1, line width=0.1pt, only marks, mark=*, mark options={solid, mycolor1}, forget plot]
  table[row sep=crcr]{%
1	1\\
2	1\\
3	1\\
4	1\\
5	1\\
6	0\\
7	1\\
8	0\\
9	0\\
10	0\\
11	0\\
12	1\\
13	1\\
14	0\\
15	0\\
16	1\\
17	0\\
18	1\\
19	1\\
20	1\\
21	0\\
22	0\\
23	0\\
24	0\\
25	0\\
26	0\\
27	0\\
28	0\\
29	0\\
30	1\\
31	1\\
32	0\\
33	1\\
34	1\\
35	1\\
36	0\\
37	0\\
38	0\\
39	0\\
40	0\\
41	0\\
42	0\\
43	0\\
44	0\\
45	0\\
46	0\\
47	0\\
48	0\\
49	0\\
50	0\\
51	0\\
52	1\\
53	1\\
54	0\\
55	0\\
56	0\\
57	0\\
58	0\\
59	0\\
60	0\\
};
\end{axis}

\end{tikzpicture}%
								\caption{Event scheduling variable $\gamma$ for simulation with $\alpha=5$.} 
								\label{fig:gamma}
							\end{figure}
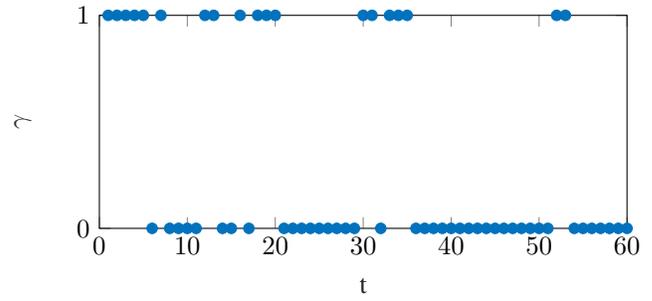

						In Figure~\ref{fig:etmhe2}, an exemplary simulation for  $\alpha=5$
						is shown. It can be observed that the trajectories of the estimated state for  $\alpha=5$ and 
						$\alpha=0$ (where  $\alpha=0$ implies that an event was triggered at every time instance, effectively functioning as a classical moving horizon estimator without event-triggering) are almost identical. 
						In  Figure \ref{fig:gamma} the corresponding values of $\gamma_t$ are plotted, i.e, the time instances when an event was scheduled are displayed.
							\par
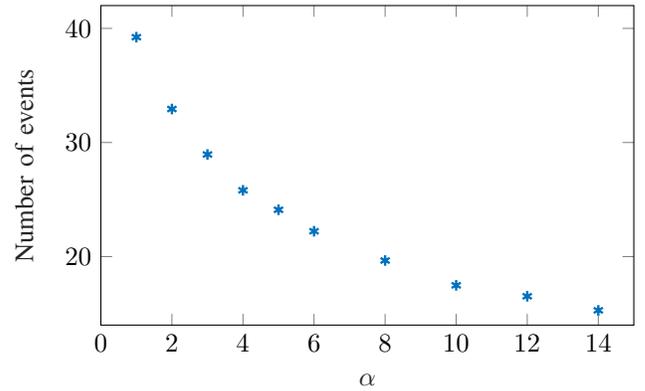
\begin{figure}[!t]
								\centering 
%
%
\definecolor{mycolor1}{rgb}{0.00000,0.44700,0.74100}%
\begin{tikzpicture}

\begin{axis}[%
width=\figurewidth,
height=0.5\figurewidth,
scale only axis,
xmin=0,
xmax=15,
xlabel style={font=\color{white!15!black}},
xlabel={$\alpha$},
ymin=14,
ymax=42,
ylabel style={font=\color{white!15!black},at={(axis description cs:0.03,0.5)}},
ylabel={Number of events},
axis background/.style={fill=white}
]
\addplot [color=mycolor1, line width=0.9pt, only marks, mark=asterisk, mark options={solid, mycolor1}, forget plot]
  table[row sep=crcr]{%
1	39.23\\
2	32.935\\
3	28.95\\
4	25.81\\
5	24.105\\
6	22.225\\
8	19.66\\
10	17.485\\
12	16.525\\
14	15.29\\
};
\end{axis}
%
\end{tikzpicture}%
									\caption{Average number of events for different choices of $\alpha$ for Example~1 (batch reactor), simulated over 60 time steps (i.e., 6 seconds).} 
									\label{fig:alphas}
								\end{figure}

 Next, we consider different values of $\alpha$ between 1 and 14. For each value of $\alpha$ we performed 200 simulations with different disturbance sequences. 
							As expected, larger values of $\alpha$  result in fewer events being triggered. This correlation is illustrated in Figure \ref{fig:alphas}.
								\color{resl}	 
							\par	After having illustrated the functionality of the proposed method, we briefly highlight some differences to the approach in \cite{Kra24} which considers a non-standard event-triggered setup that transmits a sequence of measurements  to the remote estimator upon an event.
									The ET-MHE scheme introduced in \cite{Kra24} primarily aims at reducing the number of transmission instants, whereas the approach presented in Section~\ref{sec:MHEScheme} additionally focuses on decreasing the total number of transmitted measurements to the remote estimator and avoiding large data packets compared to \cite{Kra24}.
										To further investigate this, both methods are applied to Example~1 and are tuned to achieve comparable estimation performance. 
										In particular, a horizon length of $M=34$ is used for both approaches, together with the values for  $P_2,Q,R$ and $\eta$ as specified above. The parameter $\alpha$ is adjusted accordingly to obtain similar estimation accuracy.  
										For the method in \cite{Kra24}, 
										$\alpha=15$ is used, while for the method in Section~\ref{sec:MHEScheme}, $\alpha=18$ is chosen. 
										\begin{figure}[!t]
											\begin{center}
%
%
\definecolor{mycolor1}{rgb}{0.00000,0.44700,0.74100}%
\begin{tikzpicture}

\begin{axis}[%
width=1.025\figurewidth,
height=0.8\figurewidth,
scale only axis,
xmin=0,
xmax=60,
xlabel style={font=\color{white!15!black}},
xlabel={$t$},
ymode=log,
ymin=0.00750541477964006,
ymax=10,
yminorticks=true,
ylabel style={font=\color{white!15!black}},
ylabel={$||e(t)||$},
axis background/.style={fill=white},
legend style={legend cell align=left,font=\footnotesize,align=left, at={(01,0.99)}, draw=white!15!black}
]
\addplot [color=red, line width=1.5pt]
  table[row sep=crcr]{%
0	4.5453272709454\\
1	1.0913444567506\\
2	0.245231177986421\\
3	0.133970787849527\\
4	0.0448404510160593\\
5	0.0387998654590366\\
6	0.0329538108658931\\
7	0.0286087523497835\\
8	0.041968648383371\\
9	0.0913295152161546\\
10	0.0859885847727787\\
11	0.0811743454398732\\
12	0.0762394709732184\\
13	0.0731144730126339\\
14	0.069753269846917\\
15	0.0677282535055613\\
16	0.0661178015756266\\
17	0.0651452893784795\\
18	0.0633516040806073\\
19	0.0620940472160403\\
20	0.0616277711338555\\
21	0.0613953066381119\\
22	0.0287063279151835\\
23	0.0638128589407481\\
24	0.0619197584653847\\
25	0.0612208302840857\\
26	0.0605168704114761\\
27	0.0600595822534296\\
28	0.0606057298856506\\
29	0.0610035751605689\\
30	0.0612488706650737\\
31	0.0597684225383281\\
32	0.0160116599498633\\
33	0.0173614980217699\\
34	0.00750541477964006\\
35	0.00761877590797653\\
36	0.0300119122859209\\
37	0.0290498246181829\\
38	0.0275848530356466\\
39	0.0263715761068171\\
40	0.025926826224961\\
41	0.0249314813293547\\
42	0.0250625917114805\\
43	0.0246181826986705\\
44	0.0234056528279574\\
45	0.0238133512015724\\
46	0.022618504523901\\
47	0.0219217035779303\\
48	0.0221005419230323\\
49	0.0213818080270534\\
50	0.0205809986121936\\
51	0.0203846819239065\\
52	0.144986417783109\\
53	0.0962395367728882\\
54	0.0302833668703516\\
55	0.0292697805005347\\
56	0.0274453053474819\\
57	0.0272330712652836\\
58	0.0263617596155328\\
59	0.0256656980488367\\
60	0.0243205745234016\\
};
\addlegendentry{ET-MHE sending single measurement}

\addplot [color=mycolor1, line width=1.5pt]
  table[row sep=crcr]{%
0	4.5453272709454\\
1	1.09135100877446\\
2	0.245692858587307\\
3	0.134146528045512\\
4	0.0449255410015208\\
5	0.0388708470872568\\
6	0.0330099095769448\\
7	0.0241664645818415\\
8	0.0223423606279051\\
9	0.0211436231128634\\
10	0.0197221645420047\\
11	0.0180723696966111\\
12	0.0171080056401908\\
13	0.055064751223926\\
14	0.0518570698021489\\
15	0.0500226781925239\\
16	0.048561560548772\\
17	0.0477183290280044\\
18	0.0461149842423714\\
19	0.0449027925054618\\
20	0.0445242486349239\\
21	0.0443907427019633\\
22	0.0423647406461274\\
23	0.041418389210707\\
24	0.0398241410334533\\
25	0.0393815648501228\\
26	0.038922027854657\\
27	0.0386946086437685\\
28	0.0394573440765133\\
29	0.0400496123840506\\
30	0.0404724718066679\\
31	0.0601204458907105\\
32	0.0600667871523724\\
33	0.058963077642178\\
34	0.0577485084397696\\
35	0.0576072252428489\\
36	0.0563367741710052\\
37	0.0553924935522153\\
38	0.054231891550145\\
39	0.0531536096141229\\
40	0.0531124576322717\\
41	0.052137341711732\\
42	0.0524020055689843\\
43	0.0359709288228642\\
44	0.0365259730574036\\
45	0.0354824629820606\\
46	0.036088885911625\\
47	0.0362669841927519\\
48	0.0355868880101784\\
49	0.0358974246112383\\
50	0.0363665661380416\\
51	0.036065293458862\\
52	0.0360873484297277\\
53	0.0359243431611202\\
54	0.035884665667044\\
55	0.0359784242675722\\
56	0.0367775804059147\\
57	0.0358667075097996\\
58	0.0358159834042321\\
59	0.0356865162248678\\
60	0.0362951957974534\\
};
\addlegendentry{ET-MHE sending measurement sequence \cite{Kra24}}

\end{axis}

\end{tikzpicture}%
											\end{center}
											\caption{Estimation error over time for Example~1 using ET-MHE scheme from \cite{Kra24} (blue) and ET-MHE scheme from Section~\ref{sec:MHEScheme}
											(red). } 
											\label{fig:er_vgl}
										\end{figure}
										\begin{figure}[!t]
											\begin{center}
%
%
\definecolor{mycolor1}{rgb}{0.00000,0.44700,0.74100}%
\definecolor{mycolor2}{rgb}{0.85000,0.32500,0.09800}%
\begin{tikzpicture}

\begin{axis}[%
width=1\figurewidth,
height=0.3\figurewidth,
scale only axis,
xmin=0,
xmax=60,
xlabel style={font=\color{white!15!black}},
xlabel={t},
ytick={0,1},
xlabel={$t$},
ymin=0,
ymax=1,
ylabel={$\gamma$},
axis background/.style={fill=white},
legend style={legend cell align=left, align=left, draw=white!15!black,font=\footnotesize,at={(0.52,0.8)}, anchor=north}
]
\addplot [color=mycolor2, only marks,line width=0.7pt, mark=x, mark options={solid, mycolor2}]
  table[row sep=crcr]{%
1	1\\
2	1\\
3	1\\
4	1\\
5	0\\
6	0\\
7	0\\
8	1\\
9	1\\
10	0\\
11	0\\
12	0\\
13	0\\
14	0\\
15	0\\
16	0\\
17	0\\
18	0\\
19	0\\
20	0\\
21	0\\
22	1\\
23	1\\
24	0\\
25	0\\
26	0\\
27	0\\
28	0\\
29	0\\
30	0\\
31	0\\
32	1\\
33	1\\
34	1\\
35	0\\
36	1\\
37	0\\
38	0\\
39	0\\
40	0\\
41	0\\
42	0\\
43	0\\
44	0\\
45	0\\
46	0\\
47	0\\
48	0\\
49	0\\
50	0\\
51	0\\
52	1\\
53	1\\
54	1\\
55	0\\
56	0\\
57	0\\
58	0\\
59	0\\
60	0\\
};
\addlegendentry{ET-MHE sending single measurement}

\addplot [color=mycolor1, only marks,line width=0.7pt, mark=o, mark options={solid, mycolor1}]
  table[row sep=crcr]{%
1	1\\
2	1\\
3	1\\
4	1\\
5	0\\
6	0\\
7	1\\
8	0\\
9	0\\
10	0\\
11	0\\
12	0\\
13	1\\
14	0\\
15	0\\
16	0\\
17	0\\
18	0\\
19	0\\
20	0\\
21	0\\
22	0\\
23	0\\
24	0\\
25	0\\
26	0\\
27	0\\
28	0\\
29	0\\
30	0\\
31	1\\
32	0\\
33	0\\
34	0\\
35	0\\
36	0\\
37	0\\
38	0\\
39	0\\
40	0\\
41	0\\
42	0\\
43	1\\
44	0\\
45	0\\
46	0\\
47	0\\
48	0\\
49	0\\
50	0\\
51	0\\
52	0\\
53	0\\
54	0\\
55	0\\
56	0\\
57	0\\
58	0\\
59	0\\
60	0\\
};
\addlegendentry{ET-MHE sending measurement sequence \cite{Kra24}}

\end{axis}

\end{tikzpicture}%
											\end{center}
											\caption{Event scheduling variable $\gamma$ for the experiment in Figure~\ref{fig:er_vgl}.} 
											\label{fig:trig_vgl}
										\end{figure}
										Figure~\ref{fig:er_vgl} shows the state estimation error for both methods under the same noise realization, and Figure~\ref{fig:trig_vgl} depicts the corresponding triggering instants. 
										The method from \cite{Kra24} results in 8 triggering instances, whereas the method from Section~\ref{sec:MHEScheme}
										leads to 15 events.
										However, in the method from \cite{Kra24} the sequence $\{y(j)\}_{j=\max\{t-M,\epsilon_t\}}^{t-1}$ is sent upon an event, thus  all measurements were ultimately transmitted to the remote side, with sequences of up to 18 measurements per event.
										The simulations were repeated for 100 different disturbance realizations.  Using the ET-MHE scheme from \cite{Kra24} resulted on average  in 6.5  events. The average length of the longest sequences was  23 measurements with a maximum observed length of 32 measurements. For the method in Section~\ref{sec:MHEScheme}, on average 14 events were scheduled, while the number of measurements transmitted per event are by design of the scheme always one.
										These results are summarized in Table~\ref{tab:vgl}. 
										\begin{table}[!t]
											\renewcommand{\arraystretch}{1}
											\caption{Comparison of number of events and transmitted measurements for the proposed ET-MHE scheme and the scheme in \cite{Kra24}}
											\label{tab:vgl}
											\centering
											\begin{tabular}{m{2.4cm}|m{2.2cm}|m{2.6cm}}
												& ET-MHE scheme sending $y(t-1)$ upon event& ET-MHE scheme sending $\{y(j)\}_{j=\max\{t-M,\epsilon_t\}}^{t-1}$ upon event \cite{Kra24}  \\ \noalign{\hrule height 1.2pt}
												Avg. number of events & \multicolumn{1}{c|}{14}  & \multicolumn{1}{c}{6.5} \\ \hline
												Avg. max measurement sequence length & \multicolumn{1}{c|}{1}& \multicolumn{1}{c}{23} \\ \hline
												Max measurement sequence length & \multicolumn{1}{c|}{1} & \multicolumn{1}{c}{32}  \\
											\end{tabular}	
										\end{table}
										This indicates that the preferable method depends on the application and on the motivation behind employing an  event-triggering strategy.
										While the ET-MHE from \cite{Kra24} can be beneficial when the primary communication constraint lies in the frequency of channel access rather than the amount of data transmitted per access, the ET-MHE scheme in Section~\ref{sec:MHEScheme} focuses on reducing the overall amount of data transmitted to the remote estimator and avoids the transmission of potentially long measurement sequences, which may be infeasible under limited communication bandwidth.	
										\color{black}			
								\subsection{Example 2: Robot Arm}
								
								\begin{figure}[!t]
									\centering
									\begin{tikzpicture}

										\draw[thick,->] (-0.5,0) -- (4.8,0) node[anchor=north] {$x$}; 
										\draw[thick,->] (0,-0.5) -- (0,4.8) node[anchor=west] {$y$}; %
										\def\lone{2.5} 
										\def\ltwo{2.5} %
										\def\thetaone{45} 
										\def\thetatwo{30} %
										\coordinate (Link1) at (\thetaone:\lone);

										\coordinate (Link2) at ($(Link1) + (\thetaone+\thetatwo:\ltwo)$);

										\draw[thick] (0,0) -- (Link1); %
										\draw[thick] (Link1) -- (Link2); %
										\fill (0,0) circle (2pt); %
										\fill (Link1) circle (2pt); %
										\fill (Link2) circle (2pt); %
										\draw[->] (1,0) arc[start angle=0, end angle=\thetaone, radius=1];
										\node at (1.4,0.3) {$\theta_1$};
									
										\draw[dashed] (Link1) -- ($(Link1)!-0.5!(0,0)$);
										
										\draw[->] ($(Link1)!-0.35!(0,0)$) arc[start angle={\thetaone}, end angle={\thetaone+\thetatwo}, radius=1];
										\node at ($(Link1)!-0.45!(0,0)+(-0.2,0.2)$) {$\theta_2$};
									\end{tikzpicture}
									\caption{Two-link robot arm moving in a 2D-plane.}
									\label{fig:robArm}
								\end{figure}
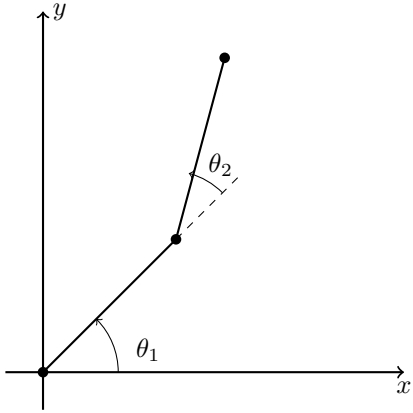
								As a second example  we consider a two-link robot arm moving in a 2D-plane as depicted in Figure~\ref{fig:robArm}. For the description of the dynamics see e.g. \cite[Example 3.2-2]{Lew04}. The system state is defined as \(x=\begin{bmatrix} \theta_1, & \theta_2, & \dot{\theta}_1, & \dot{\theta}_2 \end{bmatrix}^\top\), with the measurement output given by \(y=\begin{bmatrix} \theta_1, & \theta_2 \end{bmatrix}^\top\). Each link of the arm is assumed to have a mass of 1 kg and a length of 1 m. 
								We  applied again an Euler discretization with the sampling time $\tau=0.005$. We consider additive process and measurement noise $w=[w_1,w_2,w_3,w_4,w_5,w_6]^\top$. The initial conditions are set as \(x_0 = [\pi/4, \pi/4, 0, 0]^\top\) and \(\bar{x}_0 = [0, 0, 0, 0]^\top\). %
								For parameterizing the cost function, a quadratic i-IOSS Lyapunov function is computed following the procedure in \cite{Sch23}, with a decay rate of $\eta=0.85$, thus verifying Assumption~\ref{ass:eIOSSET}. Using condition (\ref{eq:minhor})	yields a minimum required horizon length of $M=20$  to guarantee RGES. 
								\par
\begin{figure}[!t]
											\centering
											\input{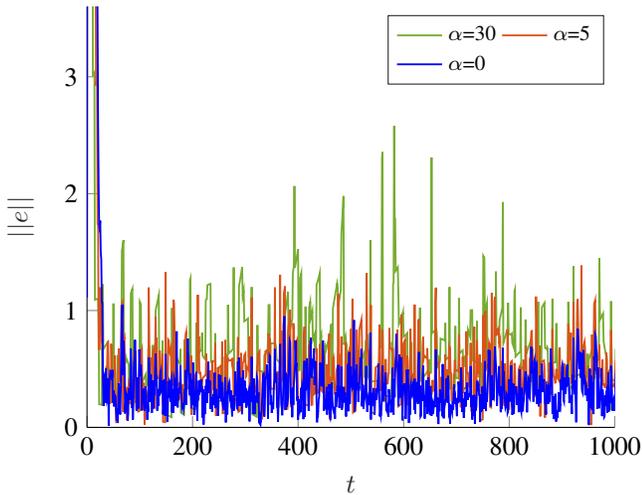}
											\caption{Estimation error over time for simulation of robot arm with different values of $\alpha$.}  
											\label{fig:e_rob_z}
										\end{figure}

								\textcolor{rev}{In the following simulation we consider a uniformly distributed process noise and measurement noise that satisfy $|w_i|\leq 0.01$, $i=1,2,3,4$ and $|w_i|\leq 0.05$, $i=5,6$, respectively. The simulation time is $5$ seconds, i.e., $N_{\text{sim}}=1000$. Figure~\ref{fig:e_rob_z} displays the estimation error for $\alpha=5$ and $\alpha=30$ as well as for $\alpha=0$. Recall that setting $\alpha=0$ yields a classical non-event-triggered moving horizon estimator. The estimation error consistently converges to a region near zero. \textcolor{rev}{Furthermore, Table~\ref{tab:er} compares the estimation error averaged over 100 simulations for $\alpha\in\{0,5,10,15,20,30,40,50,60\}$ and for two different measurement noise levels ($|w_i|\leq 0.05$ and $|w_i|\leq 0.1$, $i=5,6$).} Both Figure~\ref{fig:e_rob_z} and Table~\ref{tab:er} show that, as expected, a smaller $\alpha$, i.e., more frequent measurement transmission, results in a smaller estimation error. This behavior aligns with our theoretical findings regarding the error bound (cf.  Theorem~\ref{thm:stab2}).}				
										
\begin{figure}[!t]
											\centering
%
%
\definecolor{mycolor1}{rgb}{0.00000,0.44700,0.74100}%
\begin{tikzpicture}

\begin{axis}[%
width=\figurewidth,
height=0.5\figurewidth,
scale only axis,
xmin=0,
xmax=62,
xlabel style={font=\color{white!15!black}},
xlabel={$\alpha$},
ymin=150,
ymax=450,
ylabel style={font=\color{white!15!black}},
ylabel={Number of events},
axis background/.style={fill=white}
]
\addplot [color=mycolor1, line width=0.9pt, only marks, mark=asterisk, mark options={solid, mycolor1}, forget plot]
  table[row sep=crcr]{%
5	440.7550\\
10	335.68\\
15	287.88\\
20  259.34\\
30	226.17\\
40  207.33\\
50  193.995\\
60	184.515\\
};
\end{axis}

\end{tikzpicture}%
												\caption{Average number of events for different choices of $\alpha$ for Example~2 (robot arm), simulated over 1000 time steps (i.e., 5 seconds).} 
												\label{fig:alphasRob}
											\end{figure}
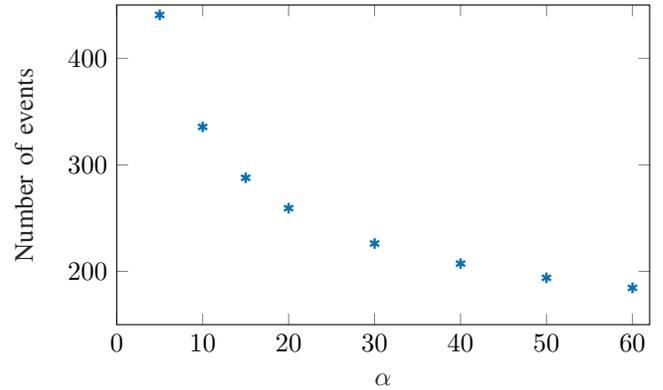

										\begin{table}
											\caption{	\color{rev}Average  estimation error for different values of $\alpha$}
											\label{tab:er}
											\centering
											\color{rev}
											\begin{tabular}{r c c}
												\hline
												\rule{0pt}{3ex} 
												\ & \multicolumn{2}{c}{
													$e_{\text{avg}}=\frac{1}{100}\frac{1}{N_{\text{sim}}}\sum_{t=1}^{N_{\text{sim}}} ||e(t)||$	
																									} 	\\ 
												\rule{0pt}{3ex} 
												$\alpha$& $|w_{i}|\leq0.05$, $i=5,6$& $|w_{i}|\leq0.1$, $i=5,6$ \\
												\hline
												\rule{0pt}{3ex}
												0&    0.5194 &   0.8340 \\[2pt]
												5&  0.6529  &  1.0884 \\[2pt]
												10   & 0.6932   & 1.1899 \\[2pt]
												15  &  0.7151   & 1.2387 \\[2pt]
												20  &  0.7320  &  1.2874 \\[2pt]
												30  &  0.7421  &  1.3337 \\[2pt]
												40  &  0.7529  &  1.3742 \\[2pt]
												50  &  0.7603 &   1.3946
												 \\[2pt]
												60   &  0.7684 &   1.4108 \\[2pt]
												\hline 
											\end{tabular}
										\end{table}
										\color{black}
										In Figure \ref{fig:alphasRob}, the average number of events  for  $\alpha\in\{5,10,15,20,30,40,50,60\}$  is presented.  
										As with the first example, an increase in $\alpha$  results in a decrease in the number of events.		
											\subsection{Additional constraint of MHE's NLP}	
											After having tested  the functionality of the algorithm with these two examples, we now examine the impact of the additional constraint (\ref{eq:NLPC4}) on the simulation results.  
											The inclusion of this constraint generally had no significant effect on the resulting estimation error in the conducted simulations. Specifically, we performed 2000 simulations for the batch reactor example (200 different disturbance sequences for each $\alpha\in\{1,2,3,4,5,6,8,10,12,14\}$). 
											For the robot arm we conducted 1600 simulations, namely 200 simulations per $\alpha\in\{5,10,15,20,30,40,50,60\}$, using 100 different disturbance sequences for each of the two measurement noise levels ($|w_i|\leq 0.1$,  $|w_i|\leq 0.05$,   $i=5,6$).
											In a substantial number of simulations, the additional constraint either did not affect the solution to the optimization problem (36 \%  (batch reactor) and 13 \% (robot arm)) or led to a slight reduction in the estimation error (49 \% (batch reactor) and 57 \%  (robot arm)).   However, as mentioned previously, even when the constraint did influence the results, the difference in estimation error remained negligibly small \color{new}(on average, the estimation error was 0.25 \%  lower when the additional constraint was used).   In the conducted simulations, the additional constraint was only active in 7.6 \% (batch reactor) and 14 \% (robot arm) of cases when the NLP was solved upon an event. \color{black}
											Therefore, one can conclude that, \color{resl}  while the additional constraint (\ref{eq:NLPC4}) was included for the benefit of our theoretical developments, its incorporation into the optimization problem is not restrictive.  \color{black} \color{new} This is in line with the earlier observation that the additional constraint is inherently satisfied by the true trajectories and is therefore reasonable to impose on the optimal solution. \color{black}		
											\subsection{Comparison with ET-MHE using a varying horizon length}
											In the following, we briefly discuss the modified algorithm presented in Section~\ref{sec:varN} that utilizes a varying horizon length. 
											We conducted again the 2000 simulations of the batch reactor and the 1600 simulations of the robot arm  as  described in the paragraph above using the ET-MHE with varying horizon length. 
											For the batch reactor, in 77\% of the simulations, the ET-MHE with a varying horizon length resulted in a smaller estimation error. The average improvement was approximately 1\%, indicating a minor benefit, though improvements were as high as 15\% in some instances. 
											In case of the robot arm, in 80\% of the simulations, the estimation error was smaller when employing the ET-MHE with a varying horizon length. On average, the improvement was about 3\%, and improvements reached up to 21\% in certain cases.
											Overall, the improvement achieved by using the ET-MHE scheme with a varying horizon length remains, on average,  relatively small.
											Nevertheless, in individual situations, we could observe a considerable advantage of this approach, as illustrated in the following example.					
											\par Figure \ref{fig:varN} illustrates a simulation in which the estimation error is significantly reduced when employing the proposed scheme with a varying horizon length. For the majority of the simulation, the estimated trajectories obtained from both the ET-MHE scheme with the fixed horizon length and the version with a varying horizon closely match. However, towards the end of the simulation, a notable deviation occurs.  At $t=54$, the estimate from the fixed-horizon scheme becomes inaccurate, whereas the varying-horizon scheme still provides a good estimate. A plausible interpretation of this behavior is that, at  $t-M=54-34=20$, a large number of consecutive events had occurred at the preceding time steps (see Figure~\ref{fig:gammafixvar}) and thus, the varying horizon scheme considered significantly more output information in  the optimization problem, enabling a more reliable estimate at this  point.			
											\begin{figure}[!t]
												\centering
													\input{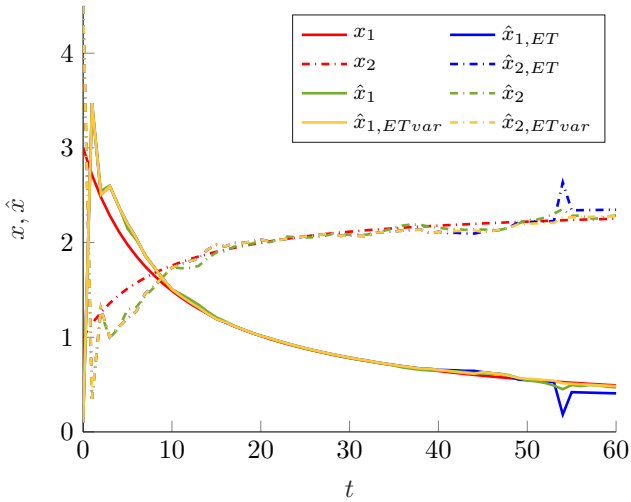}
													\caption{Comparison of ET-MHE results for $\alpha=5$ using a fixed or varying horizon length (blue and yellow, respectively), MHE estimates without event-triggering, i.e., $\alpha=0$ (green) and real system states (red).} 
													\label{fig:varN}
												\end{figure}
												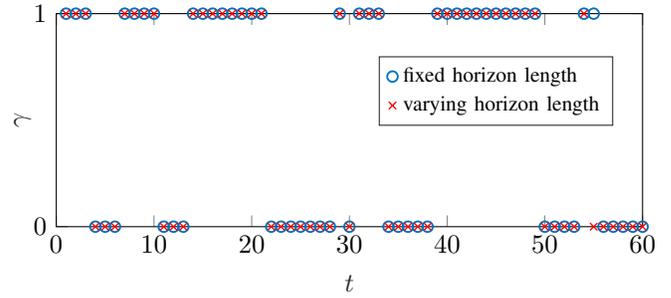
\begin{figure}[!t] 
													\begin{center}
%
%
\definecolor{mycolor1}{rgb}{0.00000,0.44700,0.74100}%
\begin{tikzpicture}

\begin{axis}[%
width=1.1\figurewidth,
height=0.3\figurewidth,
scale only axis,
xmin=0,
xmax=60,
xlabel style={font=\color{white!15!black}},
xlabel={t},
ytick={0,1},
xlabel={$t$},
ymin=0,
ymax=1,
ylabel style={font=\color{white!15!black},at={(axis description cs:0.1,0.5)}},
ylabel={$\gamma$},
axis background/.style={fill=white},
legend style={legend cell align=left, align=left, draw=white!15!black,font=\footnotesize,at={(0.75,0.8)}, anchor=north}
]
\addplot [color=mycolor1, line width=0.7pt, only marks, mark=o, mark options={solid, mycolor1}]
  table[row sep=crcr]{%
1	1\\
2	1\\
3	1\\
4	0\\
5	0\\
6	0\\
7	1\\
8	1\\
9	1\\
10	1\\
11	0\\
12	0\\
13	0\\
14	1\\
15	1\\
16	1\\
17	1\\
18	1\\
19	1\\
20	1\\
21	1\\
22	0\\
23	0\\
24	0\\
25	0\\
26	0\\
27	0\\
28	0\\
29	1\\
30	0\\
31	1\\
32	1\\
33	1\\
34	0\\
35	0\\
36	0\\
37	0\\
38	0\\
39	1\\
40	1\\
41	1\\
42	1\\
43	1\\
44	1\\
45	1\\
46	1\\
47	1\\
48	1\\
49	1\\
50	0\\
51	0\\
52	0\\
53	0\\
54	1\\
55	1\\
56	0\\
57	0\\
58	0\\
59	0\\
60	0\\
};
\addlegendentry{fixed horizon length}

\addplot [color=red, only marks, line width=0.5pt, mark=x, mark options={solid, red}]
table[row sep=crcr]{%
	1	1\\
	2	1\\
	3	1\\
	4	0\\
	5	0\\
	6	0\\
	7	1\\
	8	1\\
	9	1\\
	10	1\\
	11	0\\
	12	0\\
	13	0\\
	14	1\\
	15	1\\
	16	1\\
	17	1\\
	18	1\\
	19	1\\
	20	1\\
	21	1\\
	22	0\\
	23	0\\
	24	0\\
	25	0\\
	26	0\\
	27	0\\
	28	0\\
	29	1\\
	30	0\\
	31	1\\
	32	1\\
	33	1\\
	34	0\\
	35	0\\
	36	0\\
	37	0\\
	38	0\\
	39	1\\
	40	1\\
	41	1\\
	42	1\\
	43	1\\
	44	1\\
	45	1\\
	46	1\\
	47	1\\
	48	1\\
	49	1\\
	50	0\\
	51	0\\
	52	0\\
	53	0\\
	54	1\\
	55	0\\
	56	0\\
	57	0\\
	58	0\\
	59	0\\
	60	0\\
};
\addlegendentry{varying horizon length}

\end{axis}

\end{tikzpicture}%
													\end{center}
													\caption{Event scheduling variable $\gamma$ for the experiment in Figure~\ref{fig:varN}.} 
													\label{fig:gammafixvar}
												\end{figure}
												\color{resl}
													\subsection{Computational aspects}
												We conclude the simulation section by providing some results on the computational effort of the presented methods. For both Example~1 and Example~2, we evaluate the computation times of our main ET-MHE approach presented in Section~\ref{sec:MHEScheme} and the ET-MHE scheme  with varying horizon length of Section~\ref{sec:varN}. Moreover, we are interested in evaluating the computational cost of using the additional constraint (\ref{eq:NLPC4}) in our first approach, and hence repeat the simulations but without this constraint.	
													For each example, 300 simulations were performed using different noise realizations and values of $\alpha$, with $\alpha\in\{5,10,15\}$ for Example~1 and $\alpha\in\{10,30,60\}$ for Example~2. 
													The	simulations were performed in Matlab R2026a  using CasADi and the solver IPOPT  on an Intel Core i7-13700K.	
													The average computation time at the time steps where an event is triggered and thus the nonlinear MHE optimization problem is solved numerically is reported in Table~\ref{tab:CompTime}.
													\begin{table}[!t]
														\centering
														
														\renewcommand{\arraystretch}{1.5}
														\setlength{\tabcolsep}{8pt}
														
														\caption{Average computation time (in $s$) over event-triggered time instants.}
														\label{tab:CompTime}
														
														\begin{tabular}{m{4.2cm} c c}
															\toprule
															Method & Example 1 & Example 2 \\
															\midrule
															ET-MHE (Section~\ref{sec:MHEScheme}) & 0.0472 & 0.0160 \\
															ET-MHE without additional constraint~\eqref{eq:NLPC4} & 0.0317 & 0.0096 \\
															ET-MHE with varying horizon (Section~\ref{sec:varN}) & 0.0609 & 0.0435 \\
															\bottomrule
														\end{tabular}
													\end{table}
													As expected, adding an additional constraint to a nonlinear, nonconvex optimization problem increases the computation time. Furthermore, as already mentioned in Section~\ref{sec:varN}, using a varying horizon length is the most computationally demanding approach, due to (i) occasionally longer horizons  leading to larger optimization problems, and (ii) the need to handle a changing optimization problem structure as the horizon length varies across event times. 
										\par  To conclude, although incorporating the additional constraint (\ref{eq:NLPC4}) increases the computation time, it provides attractive theoretical guarantees for the proposed methods. Allowing for a varying horizon length further increases the computational effort but yields even stronger stability guarantees (a smaller minimum required horizon length and tighter error bounds) and may  improve performance.
													\color{black}
												\section{Conclusion}
												\label{sec:con}
												This paper presents an event-triggered moving horizon state estimator. The novel even-triggering condition and the optimization problem of the MHE are designed such that  the optimization problem only needs to be solved  if an event is scheduled and such that  only a single measurement needs to be transmitted.
												When an event occurs, the current state estimate is obtained by solving the MHE optimization problem, while open-loop predictions are used to compute state estimates between events. Additionally, we demonstrated that if the system is exponentially i-IOSS, the proposed ET-MHE achieves robust global exponential stability. Furthermore, we also showed that with the adoption of a varying horizon length, a tighter bound on the estimation error can be achieved. The effectiveness of the proposed ET-MHE scheme was illustrated through two numerical examples.  Future research could explore adapting the ET-MHE framework for networked control systems with multiple interconnected components, where the efficient sharing and distribution of limited resources become crucial. 

\section*{REFERENCES}

\def\refname{}

\begin{IEEEbiography}[{\includegraphics[width=1in,height=1.25in,clip,keepaspectratio]{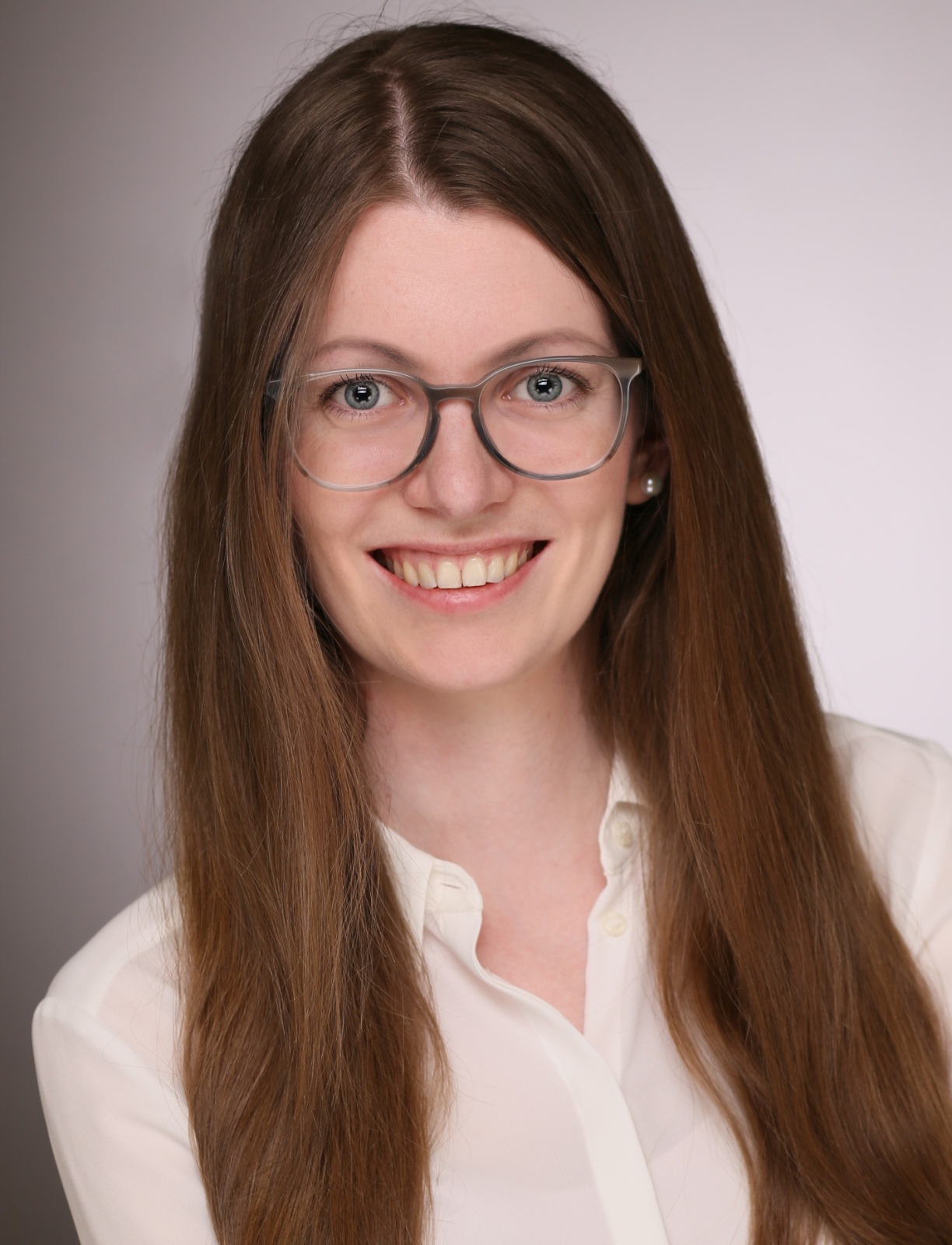}}]{Isabelle Krauss } received her Master degree
	in Automation Engineering from RWTH Aachen University, Germany, in 2020. Since 2021, she
	has been a research assistant at the Institute
	of Automatic Control, Leibniz University Hannover, Germany, where she is currently working on her
	Ph.D. under the supervision of Prof. Matthias
	A. Müller. Her research interests are in the area of observability and state estimation with focus on nonlinear systems. 
\end{IEEEbiography}
\begin{IEEEbiography}[{\includegraphics[width=1in,height=1.25in,clip,keepaspectratio]{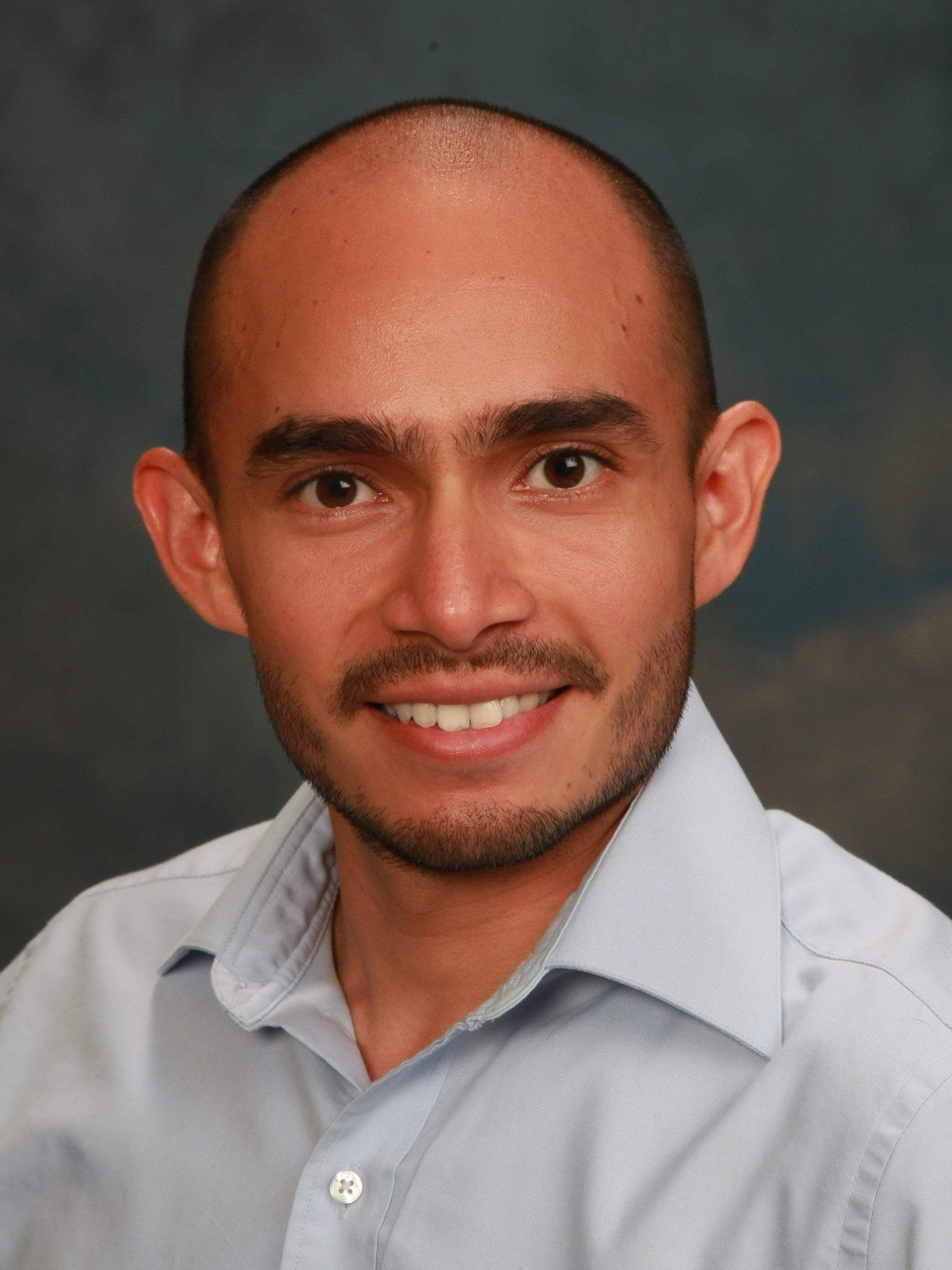}}]{Victor G. Lopez } (Member, IEEE)
	received his B.Sc. degree in Communications and Electronics Engineering from the Universidad Autonoma de Campeche, in Campeche, Mexico, in 2010, the M.Sc. degree in Electrical Engineering from the Research and Advanced Studies Center (Cinvestav), in Guadalajara, Mexico, in 2013, and his PhD degree in Electrical Engineering from the University of Texas at Arlington, Texas, USA, in 2019. In 2015 Victor was a Lecturer at the Western Technological Institute of Superior Studies (ITESO) in Guadalajara, Mexico. From August 2019 to June 2020, he was a postdoctoral researcher at the University of Texas at Arlington Research Institute and an Adjunct Professor in the Electrical Engineering department at UTA. Victor is currently a postdoctoral researcher at the Institute of Automatic Control, Leibniz University Hanover, in Hanover, Germany. His research interest include cyber-physical systems, reinforcement learning, game theory, distributed control and robust control. 
\end{IEEEbiography}
\begin{IEEEbiography}[{\includegraphics[width=1in,height=1.25in,clip,keepaspectratio]{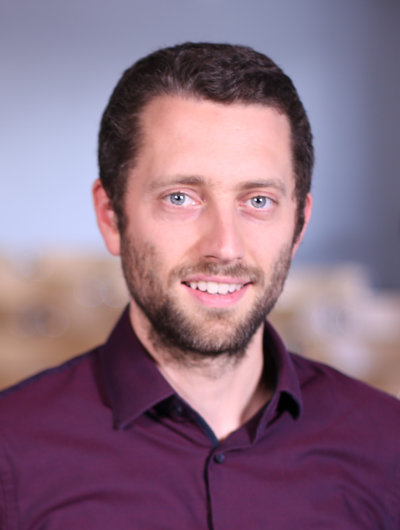}}]{Matthias A. Müller } (Senior Member, IEEE) received a Diploma degree in engineering cybernetics from 
	the University of Stuttgart, Germany, an M.Sc. in electrical and computer 
	engineering from the University of Illinois at Urbana-Champaign (both in 2009), 
	and a Ph.D. in mechanical engineering from the University of Stuttgart in 2014. 
	Since 2019, he is Director of the Institute of Automatic Control and Full Professor 
	at the Leibniz University Hannover, Germany.
	
	His research interests include nonlinear control and estimation, model predictive 
	control, and data- and learning-based control, with application in different fields 
	including biomedical engineering and robotics. He has received various distinctions for his 
	work, including the European Systems \& Control PhD Thesis Award, an ERC Starting Grant 
	from the European Research Council, the IEEE CSS George S. Axelby Outstanding Paper 
	Award, the Brockett-Willems Outstanding Paper Award, and the Journal of Process Control 
	Paper Award. He serves/d as an associate editor for Automatica and as an editor of the 
	International Journal of Robust and Nonlinear Control. 
\end{IEEEbiography}

\end{document}